\documentclass[prd,preprint]{revtex4}
\usepackage{epsfig}
\newcommand{\fms}[1]{{#1}\!\!\!/}
\newcommand{\fmsl}[1]{{#1}\!\!\!\!/}

\newcommand{\D}{\mathcal{D}}
\newcommand{\nn}{\frac{\fms{\overline{n}}}{2}} 
\newcommand{\eps}{\varepsilon}
\newcommand{\s}{\sigma_{\mu \nu}}
\newcommand{\x}{x_{\gamma}}

\begin{document}
\title{Quark mass effects in the soft-collinear effective theory and 
$\overline{B} \to X_s \gamma$ in the endpoint region}
\author{Junegone Chay}\email{chay@korea.ac.kr}
\affiliation{Department of Physics, Korea University, Seoul 136-701, Korea} 
\author{Chul Kim}\email{chk30@pitt.edu}
\affiliation{Department of
  Physics and Astronomy, University of Pittsburgh, PA 15260, U.S.A.}
\author{Adam K. Leibovich}\email{akl2@pitt.edu}
\affiliation{Department of
  Physics and Astronomy, University of Pittsburgh, PA 15260, U.S.A.}

\date{May 5, 2005\\ \vspace{1cm} }

\baselineskip 3.0ex \parskip 2.0ex

\begin{abstract}
\baselineskip 3.0ex 
We consider the effects of a light quark mass in the soft-collinear
effective theory (SCET) and we apply them to $\bar{B} \to X_s
\gamma$ in the endpoint region. We find that the reparameterization
invariance can be extended by including the collinear quark  mass in
the SCET Lagrangian. This symmetry constrains the theory with the
quark mass terms, and we present explicit results at one loop. It
also relates the Wilson coefficients of some mass operators to those
of the leading operators, which are useful in organizing the
subleading effects due to the quark mass in $\bar{B} \to X_s \gamma$.  
We present strange quark mass corrections to $\bar{B} \to X_s \gamma$
in the endpoint region as an application. The forward scattering  
amplitude from the mass corrections is factorized, and it can be
expressed as a convolution of the $m_s^2/p_X^2$-suppressed jet function 
and the leading-order shape function of the $B$ meson. This
contribution should be added to the existing subleading contributions
from the $B$ meson shape functions to obtain complete subleading
corrections.
\end{abstract}

\maketitle

\section{Introduction}
The soft-collinear effective theory (SCET) 
\cite{Bauer:2000ew,Bauer:2000yr,Bauer:2001yt} 
has been widely used to describe high-energy processes which include
energetic light particles. It is obtained from QCD by integrating out
the degrees of freedom which are larger than a typical energy
scale,  $Q$. The effective theory contains a rich class of
symmetries, and these symmetries of SCET provide us with new insight into
factorization theorems 
\cite{Bauer:2001yt,Bauer:2001cu,Bauer:2002nz} 
and enable us to perform a systematic power counting in hadronic
processes \cite{Powercounting}. SCET has been applied successfully to
many high energy processes such as exclusive $B$ decays 
\cite{Chay:2002vy, Htl, Leptonic, Nonleptonic,
  Mantry:2003uz,Chay:2003kb}, inclusive $B$ 
decays~\cite{Bauer:2000ew,Inclusive}, quarkonium production and
decay~\cite{Quarkonium}, deep inelastic
scattering~\cite{Manohar:2003vb}, and jet physics~\cite{Jet}.

In SCET, the momentum of a light energetic particle has three distinct
scales  and can be written as   
\begin{equation}
p^{\mu} = \overline{n}\cdot p \frac{n^{\mu}}{2} +p_{\perp}^{\mu}
+n\cdot p \frac{\overline{n}^{\mu}}{2} = \mathcal{O} (\lambda^0) +
\mathcal{O} (\lambda^1) +\mathcal{O} (\lambda^2).
\end{equation}
Here $n$ and $\overline{n}$ are light-cone vectors satisfying 
$n^2=\overline{n}^2=0,~n\cdot \overline{n} =2$ and  
$\lambda$ is a small parameter. In many processes, $\lambda$ is chosen as 
$\sqrt{\Lambda/Q}$ or $\Lambda/Q$, where $\Lambda$ is a typical
hadronic scale. The effective theory which has
a small expansion parameter $\lambda \sim  
\sqrt{\Lambda/Q}$ is called $\mathrm{SCET}_{\mathrm{I}}$ and 
the effective theory in which physical quantities are expanded in
powers of $\lambda \sim \Lambda/Q$ is called
$\mathrm{SCET}_{\mathrm{II}}$. If there are contributions at
intermediate scales of order $\sqrt{Q\Lambda}$, we employ the two-step
matching in which $\mathrm{SCET}_{\mathrm{I}}$ is obtained from the
full theory by integrating out the hard modes of order $p^2 \sim Q^2$,
and $\mathrm{SCET}_{\mathrm{II}}$ is obtained by successively
integrating out hard-collinear modes of order $p^2 \sim Q \Lambda$
\cite{Bauer:2001yt}.  

At leading order in SCET, the collinear quarks are regarded as
massless. Because the mass of a light quark, $m$, is very small 
compared to the hard scale $Q$ or the intermediate hard-collinear scale
$\sqrt{Q \Lambda}$, the quark mass can be neglected at leading order
in $\lambda$. However, the light-quark mass terms
\cite{Chay:2003kb,Rothstein:2003wh,Leibovich:2003jd,Chen:2003fp}, and
in some situations the charm quark mass \cite{Boos:2005by} can be included in
the framework of SCET.
In Ref.~\cite{Leibovich:2003jd}, the authors first considered the
quark mass in the SCET Lagrangian. Any operators including the light
quark mass are formally suppressed by $\Lambda/Q$ or more compared to
the leading contribution. However if there are no leading terms, the
quark mass can appear at leading order. $SU(3)$ breaking effects
can be of this type since the strange quark mass can be numerically
regarded as of order $\Lambda$ (it is not possible to treat isospin
breaking effects in this way since the masses of the up and down
quarks are too small to be regarded as of order $\Lambda$). Another
remarkable point about the quark mass is that it can give an 
enhanced contribution to some hadronic processes in
$\mathrm{SCET}_{\mathrm{II}}$ due to the different power counting schemes
in $\mathrm{SCET}_{\mathrm{I}}$ and $\mathrm{SCET}_{\mathrm{II}}$. 
Although they do not appear at leading order in
$\mathrm{SCET}_{\mathrm{I}}$, since the quark mass terms
are suppressed by $\Lambda/Q$, light quark masses can give significant
corrections to the matching process related to hard-collinear degrees of
freedom. The contribution of the quark mass to the decay rate can be of
order $m^2/(Q^2(1-x)) \sim \Lambda /Q$ near the endpoint $1-x \sim
\Lambda/Q$. This is one of the main themes to be investigated in this
paper.  
  
SCET can be extended to include the light quark mass, which we
regard as of order $\Lambda$. We can systematically implement the
quark mass in SCET and consider its renormalization behavior. We
find that the reparameterization invariance
\cite{Chay:2002vy,Manohar:2002fd} still holds for the transformations
of type-I and type-III in spite of the presence of the quark mass. But
the transformation of type-II does not hold in its original form. However the
transformation of type-II can be modified (or extended) to 
include the quark mass so that the symmetry still exists. This
extended reparameterization invariance relates the leading operators to
some subleading operators that include the quark mass. In practical
applications, the strange quark mass is the only light quark mass that is 
relevant and we consider the
quark mass effects in $\bar{B} \to X_s \gamma$ near the
endpoint as a concrete example. Naively, the mass terms give corrections of order
$m^2/m_b^2$ compared to the leading order contribution. But
contributions of order $m^2/[m_b^2 (1-\x)]$
with $\x= 2E_{\gamma}/m_b$ can arise, which are of 
order $\Lambda/m_b$ near the
endpoint region. 

In this paper we investigate the effects of the quark mass 
in SCET  and consider the symmetries including a
quark mass. We also consider the renormalization effects and 
the Wilson coefficients of the mass operators in SCET.
We then apply these results to $\bar{B} \to X_s \gamma$
in the endpoint region and discuss the contribution of the quark mass terms. 
In section II, the SCET Lagrangian with the light quark mass is
constructed.  We find an extended reparameterization transformation 
under which the Lagrangian is invariant, and we divide the
Lagrangian into two reparameterization-invariant combinations. 
In this procedure, we show that the original reparameterization
invariance symmetry without a quark mass can be extended by modifying
the transformation of the collinear quark. We
describe the consequence of the extended reparameterization invariance
on the renormalization behavior of the mass operators.  
In section III, the Wilson coefficients of the effective operators
including the quark mass are obtained to first order in $\alpha_s$
from the matching between full QCD and SCET. Also their
renormalization behavior is presented with the effective theory quark mass 
renormalization at one loop. In section IV, the corrections due to
the strange quark mass in $\overline{B}\rightarrow X_s \gamma$ near
the endpoint region are considered. They can give corrections of order
$\Lambda/m_b$, contrary to naive expectations. From the matching 
of the heavy-to-light current between the full theory and $\rm{SCET_I}$,  
we obtain the subleading current operators including the quark mass.
We then consider the time-ordered products of  the 
currents and mass operators contributing to the decay rate in SCET. 
We show that the forward scattering amplitude with the mass
corrections factorizes, similar to the leading-order result, and
the jet function can be expanded in powers 
of the quark mass. Finally the results are summarized and the conclusions
are presented in the final section.

\section{Mass operators and the reparameterization invariance}
In SCET, the collinear quark in the full theory is decomposed into
\begin{eqnarray} 
\psi(x)&=&\sum_{\tilde{p}} e^{-i\tilde{p} \cdot x} q_{n,p}(x)  
        = \sum_{\tilde{p}} e^{-i\tilde{p} \cdot x} 
         \Bigl(\frac{\fms{n}\fms{\overline{n}}}{4}+
               \frac{\fms{\overline{n}}\fms{n}}{4} \Bigr)q_{n,p}(x)
\nonumber \\ 
&=&\sum_{\tilde{p}} e^{-i\tilde{p} \cdot x} 
 \Bigl(\xi_{n,p} (x) +\xi_{\overline{n},p} (x)\Bigr),
\label{quark}
\end{eqnarray} 
where $\tilde{p}^{\mu}$ is a label momentum, and 
$\frac{\fms{n}\fms{\overline{n}}}{4} q_{n,p} = \xi_{n,p}$,
$\frac{\fms{\overline{n}}\fms{n}}{4} q_{n,p} = \xi_{\overline{n},p}$
are the projected spinors. After integrating out the off-shell field
$\xi_{\bar{n},p}$ \cite{Leibovich:2003jd}, the SCET Lagrangian with a
quark mass is written as
\begin{eqnarray} 
 \mathcal{L}_{\mathrm{SCET}} &=& \overline{\xi}_{n,p'} n\cdot i\D 
\frac{\fms{\overline{n}}}{2} \xi_{n,p}
+ \overline{\xi}_{n,p'} i\fmsl{\D}_{\perp} 
\frac{1}{\overline{n}\cdot i\D} i\fmsl{\D}_{\perp} 
\frac{\fms{\overline{n}}}{2} \xi_{n,p}
\nonumber \ \\
&& +~ m\overline{\xi}_{n,p'} \bigl[i \fmsl{\D}_{\perp},~ 
  \frac{1}{\overline{n}\cdot i\D}\bigr] \frac{\fms{\overline{n}}}{2} \xi_{n,p}
 - m^2 \overline{\xi}_{n,p'} \frac{1}{\overline{n}\cdot i\D} 
\frac{\fms{\overline{n}}}{2} \xi_{n,p},
\label{L1}
\end{eqnarray} 
where a summation over the label momenta is implied, and the covariant
derivative $\mathcal{D}^{\mu}$ is given by \cite{Bauer:2000yr}
\begin{eqnarray}
&&\D^{\mu}=D^{\mu}_c + D^{\mu}_{us}, \nonumber \\ 
&& iD^{\mu}_c = \mathcal{P}^{\mu} + g A^{\mu}_{n,q},~~~
   iD^{\mu}_{us} = i \partial^{\mu} + g A^{\mu}_{us}.
\end{eqnarray}

Let us consider first $\mathrm{SCET}_{\mathrm{I}}$ with the expansion
parameter $\lambda \sim \sqrt{\Lambda/Q}$, in which the ultrasoft
(usoft) fields can interact with the collinear fields. The usoft
momentum is of order $\Lambda$, and $p_{\perp} \sim
\sqrt{Q\Lambda}$. For a collinear strange quark, if we treat the sizes 
of the quark mass $m$ and $iD_{us}$ to be of the same order
$O(\lambda^2)$, the term proportional to $m$ in Eq.~(\ref{L1}) is of
order $\mathcal{O}(\lambda)$ and the term proportional to $m^2$ starts from 
$\mathcal{O}(\lambda^2)$.  In this case, the mass terms in SCET are 
suppressed at least by order $\lambda$ compared to the leading
Lagrangian, and the spin of the collinear quark is preserved at
leading order in SCET.  

Integrating out the hard-collinear degrees of freedom with $p^2_{hc} \sim
Q\Lambda$ to obtain $\mathrm{SCET_{II}}$, the usoft fields are
decoupled from the collinear fields \cite{Bauer:2001yt}, and 
the Lagrangian of the collinear quark sector in $\mathrm{SCET_{II}}$
can be written as 
\begin{eqnarray}  
 \mathcal{L}_c^{\mathrm{II}} &=& \overline{\xi}_{n,p'} n\cdot iD_c 
\frac{\fms{\overline{n}}}{2} \xi_{n,p}
+ \overline{\xi}_{n,p'} i\fmsl{D}^{\perp}_c 
\frac{1}{\overline{n}\cdot iD_c} i\fmsl{D}^{\perp}_c 
\frac{\fms{\overline{n}}}{2} \xi_{n,p}
\nonumber \ \\
&& +~ m\overline{\xi}_{n,p'} \bigl[i \fmsl{D}^{\perp}_c,~ 
  \frac{1}{\overline{n}\cdot iD_c}\bigr] \frac{\fms{\overline{n}}}{2}
  \xi_{n,p}  - m^2 \overline{\xi}_{n,p'} \frac{1}{\overline{n}\cdot
    iD_c}  \frac{\fms{\overline{n}}}{2} \xi_{n,p},
\label{L2}
\end{eqnarray} 
where 
\begin{eqnarray} 
i D^{\mu}_c&=&\bigl(
\overline{n} \cdot \mathcal{P} + g \overline{n} \cdot A_{n,q}\bigr)
\frac{n^{\mu}}{2} + 
 \bigl(\mathcal{P}^{\mu}_{\perp} + g A^{\mu}_{n,q,\perp} \bigr) + 
\bigl(n \cdot \mathcal{P}+ g n \cdot A_{n,q} \bigr) 
\frac{\overline{n}^{\mu}}{2} \nonumber \\
&=&\mathcal{O}(\lambda^0) + \mathcal{O}(\lambda^1) +\mathcal{O}(\lambda^2). 
\end{eqnarray} 
Here the expansion parameter $\lambda$ is of order $\Lambda/Q$, and
the collinear fields have momenta  $p_c^2 \sim \Lambda^2$.  
Contrary to $\mathrm{SCET}_{\mathrm{I}}$, the mass terms in
Eq.~(\ref{L2}) belong to the leading-order Lagrangian. Therefore the
effects of the mass terms can be important at leading order.

Before we investigate the effects of the radiative corrections for the
new operators with a quark mass in Eq.~(\ref{L2}), it is useful to
consider the symmetries of SCET with the quark mass. 
In Refs.~\cite{Chay:2002vy,Manohar:2002fd}, it has been 
shown that the SCET Lagrangian without the quark mass has a
reparameterization invariance. One of the consequences is that the
kinetic energy in SCET is not renormalized to all orders in
$\alpha_s$. And when we consider current operators in SCET, there are
subleading operators which form a reparameterization-invariant
combination with the leading operators. In this case, the Wilson
coefficients of these subleading operators are the same as those of
the leading operators to all orders in $\alpha_s$.  
When the mass terms are included in SCET, the situation is slightly
different. In this case, we can find an extended reparameterization
transformation under which the Lagrangian is still invariant, and the
Lagrangian consists of two independent sets of the 
operators which are separately reparameterization invariant. A similar
example exists in the heavy quark effective theory (HQET) 
\cite{Eichten:1990vp,Falk:1990pz,Amoros:1997rx},  
in which the chromomagnetic operator belongs to a different
reparameterization invariant combination
from the kinetic term in HQET, and has a nontrivial Wilson
coefficient.

Let us consider the effect of the mass term on the reparameterization
invariance and how we can extend the reparameterization symmetry with
the quark mass. The Lagrangian before integrating out
$\xi_{\bar{n},p}$ is given by    
\begin{equation} 
\mathcal{L} = \sum_{\tilde{p},\tilde{p'}} e^{i (\tilde{p'}
  -\tilde{p})\cdot x} \Bigr(\overline{q}_{n,p'} i \fmsl{\D}~ q_{n,p} -
m~\overline{q}_{n,p'} q_{n,p}  \Bigl),
\label{Lr}
\end{equation} 
where the quark field in SCET is given by
$q_{n,p}=\xi_{n,p}+\xi_{\bar{n},p}$, and the covariant derivative
is $\mathcal{D}^{\mu}=D^{\mu}_c + D^{\mu}_{us}$. Here
the covariant derivative $\mathcal{D}^{\mu}$ is invariant under the
reparameterization transformation since it is a four-vector, which
does not change under a different basis of $n^{\mu}$ and
$\overline{n}^{\mu}$. Furthermore, the quantity $\sum_{\tilde{p}}e^{-i
\tilde{p} \cdot x} q_{n,p}$ is the quark field in the full theory,
which also does not change under the reparameterization
transformation. Therefore the two terms in Eq.~(\ref{Lr}) are separately 
reparameterization invariant.  Thus, there are two
independent reparameterization-invariant combinations in
Eq.~(\ref{L1}), (\ref{L2}), and (\ref{Lr}) if we can still find the
appropriate reparameterization invariance. 

In fact, there is a reparameterization invariance which can be
extended to the case with the mass term. The original
reparameterization invariance combined with the gauge invariance
requires that the covariant derivative $\D_{\mu}$ not 
change under the transformations of type-I, II and III in
Ref.~\cite{Manohar:2002fd}. We can find the same types of the
reparameterization transformations under which the Lagrangian with the
quark mass is invariant. In this case, we only need to check 
if the quark field $\sum_{\tilde{p}}e^{-i \tilde{p} \cdot
  x} q_{n,p}$ in the full theory remains invariant under these three
types of the transformation. Using the equation of motion, we can
write the quark field in the full theory in terms of $\xi_{n,p}$ as 
\cite{Leibovich:2003jd} 
\begin{equation}
\psi (x) = \sum_{\tilde{p}}e^{-i \tilde{p} \cdot x} q_{n,p}  = 
 \sum_{\tilde{p}}e^{-i \tilde{p} \cdot x} \Bigl[1 + 
 \frac{1}{\overline{n} \cdot i\mathcal{D}} \bigl(i \fmsl{\D}_{\perp} + m
 \bigr) \frac{\fms{\overline{n}}}{2} \Bigr] \xi_{n,p}.
\label{psi}
\end{equation}
Without the mass term, the quark field $\psi$ has the original
reparameterization invariance. 

With the mass term, $\psi$ is not invariant
under all of the original reparameterization transformations. $\psi$ is still 
invariant under the
reparameterization transformations of type-I and III,
but not the transformation of type-II.
In order to see this, it is enough to look at the term proportional to
the quark mass in Eq.~(\ref{psi}) under the type-II transformation,
in which the light-cone vector $\overline{n}^{\mu}$ changes to
$\overline{n}^{\mu} + \varepsilon^{\mu}_{\perp}$ with infinitesimal
$\varepsilon^{\mu}_{\perp}$. The transformation yields    
\begin{eqnarray}  \label{ftwo}
 \frac{m}{\overline{n} \cdot i\mathcal{D}}
 \frac{\fms{\overline{n}}}{2}  \xi_{n}   &\rightarrow&
 \frac{m}{\overline{n} \cdot i\D + \varepsilon_{\perp} \cdot i \D_{\perp}}
 \Bigl(\nn + \frac{\fms{\varepsilon}_{\perp}}{2} \Bigr) 
\Bigl(1+ \frac{\fms{\varepsilon}_{\perp}}{2}
 \frac{1}{\overline{n} \cdot i\D} i \fmsl{\D}_{\perp} \Bigr)\xi_{n} 
\nonumber \\ 
&=& m\Bigl( \frac{1}{\overline{n} \cdot i\mathcal{D}}
 \frac{\fms{\overline{n}}}{2} \xi_{n} 
 - \frac{1}{\overline{n} \cdot i\mathcal{D}} 
  \varepsilon_{\perp} \cdot i\D_{\perp} 
 \frac{1}{\overline{n} \cdot i\mathcal{D}} \nn \xi_{n} 
 + \frac{1}{\overline{n} \cdot i\mathcal{D}} 
 \frac{\fms{\varepsilon}_{\perp}}{2} \xi_{n}  \nonumber \\
&&+ \frac{1}{\overline{n} \cdot i\mathcal{D}} 
 \frac{\fms{\varepsilon}_{\perp}}{2} 
 \frac{1}{\overline{n} \cdot i\mathcal{D}} 
  i \fmsl{\D}_{\perp} \nn \xi_{n} \Bigr) \neq 
 \frac{m}{\overline{n} \cdot i\mathcal{D}}
 \frac{\fms{\overline{n}}}{2} \xi_{n},  
\end{eqnarray}  
which clearly shows that the mass term is not invariant under the
transformation of type-II. However, we can find an extended
transformation of the spinor under the type-II transformation such
that $\psi$ remains invariant, and in the limit of the zero quark mass,
the transformation reduces to the original transformation of type-II.

Suppose that the spinor $\xi_n$ changes as $\xi_n \rightarrow \xi_n
+\delta \xi_n$ under the transformation of type-II. Then $\psi$
transforms as
\begin{eqnarray} \label{a}
\Bigl[1 + 
 \frac{1}{\overline{n} \cdot i\mathcal{D}} \bigl(i \fmsl{\D}_{\perp} + m
 \bigr) \frac{\fms{\overline{n}}}{2} \Bigr] ~\xi_{n}~~
&\rightarrow& \Bigl[1+
\frac{1}{\overline{n} \cdot i\D + \varepsilon_{\perp} \cdot i \D_{\perp}}
\bigl( i \fmsl{\D}_{\perp} - \frac{\fms{\varepsilon}_{\perp}}{2} n\cdot i\D
- \frac{\fms{n}}{2} \varepsilon_{\perp}\cdot i\D_{\perp} + m
 \bigr)\Bigr] \nonumber \\
 &\times& \bigl(\nn + \frac{\fms{\varepsilon}_{\perp}}{2} \bigr)
  \bigl( \xi_{n} + \delta \xi_{n} \bigr). 
\end{eqnarray}
Requiring that it be invariant under the transformation,
the solution for $\delta \xi_n$ is given by
\begin{equation}
\delta \xi_n = \frac{\fms{\varepsilon}_{\perp}}{2} 
\frac{1}{\overline{n} \cdot i\mathcal{D}} \bigl(i \fmsl{\D}_{\perp} - m 
\bigr)~ \xi_n,
\label{b}
\end{equation}
which reduces to the original reparameterization transformation of
type-II without the quark mass. If we plug this solution into
Eq.~(\ref{a}), we obtain
\begin{eqnarray} 
 && \delta_{\rm{II}} \Bigl[1 + 
 \frac{1}{\overline{n} \cdot i\mathcal{D}} \bigl(i \fmsl{\D}_{\perp} + m
 \bigr) \frac{\fms{\overline{n}}}{2} \Bigr] ~\xi_{n} 
 = - \frac{1}{\overline{n} \cdot i\D} 
  \frac{\fms{\varepsilon}^{\perp}}{2} \nonumber \\  
\label{c} 
&&\times \Bigl[ n\cdot i\D + i \fmsl{\D}_{\perp} 
 \frac{1}{\overline{n} \cdot i\mathcal{D}} i \fmsl{\D}_{\perp} + m \bigl(
 i\fmsl{\D}_{\perp} \frac{1}{\overline{n} \cdot i\D} - 
 \frac{1}{\overline{n} \cdot i\D} i \fmsl{\D}_{\perp} \bigr)
 - m^2 \frac{1}{\overline{n} \cdot i\D} \Bigr]~\nn \xi_n =0, 
\end{eqnarray} 
using the equation of motion, Eq.~(\ref{L1}). So the extended
transformation of type-II on the spinor with the quark mass can be
written as  
\begin{equation} 
\xi_n ~\stackrel{\rm II}{\longrightarrow}~ 
\Bigl[~1+ \frac{\fms{\varepsilon}^{\perp}}{2} 
\frac{1}{\overline{n} \cdot i\mathcal{D}} \bigl(i \fmsl{\D}_{\perp} -
m \bigr) ~\Bigr] \xi_n.
\end{equation} 
Therefore the reparameterization symmetries in the presence of the
light quark mass in SCET still exist with the only modification of 
the spinor under the transformation of type-II, while the other
transformations remain intact. 

As mentioned above, there are two independent
reparameterization-invariant combinations in Eq.~(\ref{Lr}).  
Putting Eq.~(\ref{psi}) into Eq.~(\ref{Lr}), each combination
can be written as 
\begin{eqnarray}
\overline{q}_{n,p'} i \fmsl{\D}~ q_{n,p}&=& 
\overline{\xi}_{n,p'} \Bigl[~n\cdot i\D 
+  i\fmsl{\D}_{\perp} 
\frac{1}{\overline{n}\cdot i\D} i\fmsl{\D}_{\perp}  ~\Bigr]
\nn \xi_{n,p} + m^2~\overline{\xi}_{n,p'} \frac{1}{\overline{n}\cdot i\D} 
\frac{\fms{\overline{n}}}{2} \xi_{n,p} \nonumber \\
\label{comb1}&\equiv& \mathcal{K} - \mathcal{O}_m^{(2)}, \\
- m~\overline{q}_{n,p'} q_{n,p}&=& 
 m~\overline{\xi}_{n,p'} \bigl[i \fmsl{\D}_{\perp},~ 
  \frac{1}{\overline{n}\cdot i\D}\bigr] \frac{\fms{\overline{n}}}{2} \xi_{n,p}
 -2 m^2 \overline{\xi}_{n,p'} \frac{1}{\overline{n}\cdot i\D} 
\frac{\fms{\overline{n}}}{2} \xi_{n,p} \nonumber \\
\label{comb2}&\equiv& \mathcal{O}_m^{(1)} + 2\mathcal{O}_m^{(2)}, 
\end{eqnarray}
where $\mathcal{K}$ is the kinetic term of SCET and the mass operators 
$\mathcal{O}_m^{(i)}$ are suppressed by $\lambda^i$ compared to
$\mathcal{K}$ in $\mathrm{SCET_I}$.
Because the kinetic term in the effective theory is not renormalized
to all orders in $\alpha_s$, it is also true for the
reparameterization-invariant combination $\mathcal{K}
-\mathcal{O}_m^{(2)}$. But the other combination in Eq.~(\ref{comb2}) 
does not have  
such a constraint, and in general  
it can have a nontrivial Wilson coefficient at higher orders. 
Putting these together, to all orders in $\alpha_s$, the SCET
Lagrangian can be written as  
\begin{eqnarray}
\mathcal{L}_{\rm{SCET}}&=&\mathcal{K}-\mathcal{O}_m^{(2)}
+ C (\mu) \bigl(\mathcal{O}_m^{(1)}+2\mathcal{O}_m^{(2)}\bigr) \nonumber \\ 
&=& \mathcal{K} + C(\mu) \mathcal{O}_m^{(1)}   
 + \Bigl(-1+ 2C(\mu)\Bigr) \mathcal{O}_m^{(2)}.
\end{eqnarray}

The Wilson coefficient $C(\mu)$ can be obtained from matching the
full QCD Lagrangian onto SCET by treating the mass term as a
perturbation. As will be explicitly shown in the next section, when  
dimensional regularization is used  both for the ultraviolet and
the infrared divergences, all the radiative corrections at order
$\alpha_s$ are zero since the ultraviolet divergences cancel the
infrared divergences. Therefore there is no finite contribution in
matching, and the Wilson coefficient remains as 1. 
The SCET Lagrangian, at least to first order in $\alpha_s$, can
be written as  
\begin{equation}
\mathcal{L}_{\mathrm{SCET}} = \mathcal{K} +\mathcal{O}_m^{(1)}
+\mathcal{O}_m^{(2)}.   
\end{equation}
If the
radiative corrections remain zero at higher orders, the Wilson coefficient is
equal to 1 to all order in $\alpha_s$. An argument to the
non-renormalization to all orders was presented in the first reference
of \cite{Htl}, and in Ref.~\cite{Beneke:2003pa} including the quark
mass. 

The scaling behavior of the quark mass can be considered by extracting 
the ultraviolet divergent part in the radiative corrections of the
operators $\mathcal{O}_m^{(1,2)}$ since these operators involve the
quark mass. It can be obtained by computing the radiative corrections
for the quark mass with the wavefunction renormalization of the spinor
$\xi_n$. Physically, the scaling behavior of the quark mass should be
the same as that in the full theory since there are no degrees of
freedom integrated out, which contribute to the evolution of the quark
mass of order $\Lambda$. For example, the self energy for $\xi_n$ is
the same as that for the spinor $\psi$ in the full theory. This is in
contrast to HQET, where the magnetic operator has a nontrivial Wilson
coefficient because the hard calculation of the full theory has a
dependence on the heavy quark mass. All these aspects will be verified
explicitly to order $\alpha_s$ in the next section.

\section{Matching and renormalization of the mass operators}
The matching between full QCD and SCET can be performed by
considering the quark propagator. The quark propagator in the full
theory can be written to all orders in $\alpha_s$ as  
\begin{equation} 
\frac{i}{\fms{p}-m}
\longrightarrow \frac{i}{\fms{p}-m-\sum(\fms{p},m)}, 
\label{pro}
\end{equation} 
where $\sum(\fms{p},m)$ is the self energy of the quark,
and the higher-order corrections of the full QCD Lagrangian can be
obtained by replacing the Lagrangian in momentum space as   
\begin{equation}
\overline{\psi}~ (\fms{p}-m) ~\psi
\longrightarrow  
\overline{\psi} \Bigl[\fms{p}-m -\sum(\fms{p},m) \Bigr] \psi.
\label{lag}
\end{equation}
When we match $\rm{SCET_I}$ onto the full theory at the scale 
$\mu \sim Q$ where $Q$ is the large momentum of the collinear quark, 
the self energy can be written as 
\begin{equation}
\sum(\fms{p},m) = A(p^2,\mu)\fms{p} + B(p^2,\mu) m, 
\end{equation} 
where the virtuality of the collinear quark $p^2$ is treated as 
$\mu^2 \gg p^2 \gg m^2$.
At first order in $\alpha_s$, the coefficients are given as
\begin{eqnarray}  \label{ab}
A(p^2,\mu)&=&-\frac{\alpha_s C_F}{4\pi}  \Bigl(\frac{1}{\eps} + 
\ln \frac{\mu^2}{-p^2}+1 \Bigr), \nonumber \\
B(p^2,\mu)&=&\frac{\alpha_s C_F}{4\pi} \Bigl(\frac{4}{\eps} + 
4\ln \frac{\mu^2}{-p^2}+6 \Bigr), 
\end{eqnarray}
where $D=4-2\eps$ and $1/\eps$ represents the ultraviolet divergence
and the infrared divergences are regulated by the logarithmic
terms. This method is useful in extracting the ultraviolet
divergences. For example, the counterterms for the wavefunction
renormalization $Z_{\psi}$ and the mass renormalization $Z_m$ are
given by 
\begin{equation}
Z_{\psi} =    1- \frac{\alpha_s C_F}{4\pi} \frac{1}{\eps}, \ Z_m =  1-
\frac{\alpha_s C_F}{4\pi} \frac{3}{\eps}.   
\end{equation}

A more convenient method is to use pure dimensional regularization
with all the external particles on their mass shell. This greatly
simplifies the computation both in the full theory and in the
effective theory. In both theories the on-shell graphs have no finite
parts since there are scaleless integrals, which vanish in pure
dimensional regularization. Furthermore the matching results are gauge
independent and renormalization-scheme independent only when we put
the external particles on their mass shell. Eq.~(\ref{ab}) can be
written in pure dimensional regularization as
\begin{equation}
A(\mu) =-\frac{\alpha_s C_F}{4\pi}  \Bigl(\frac{1}{\eps}
-\frac{1}{\varepsilon_{\mathrm{IR}}} \Bigr), \ 
B(\mu)=\frac{\alpha_s C_F}{4\pi} \Bigl(\frac{4}{\eps}
-\frac{4}{\varepsilon_{\mathrm{IR}}} \Bigr),  
\end{equation}
where the infrared poles in $\varepsilon_{\mathrm{IR}}$ can be
explicitly computed or can be inferred from the ultraviolet divergence
with the fact that the radiative corrections are zero. 

At one loop after the ultraviolet divergence is removed, the radiative
correction of the full QCD Lagrangian is given by
\begin{equation} 
\Bigl(1-\frac{\alpha_s C_F}{4\pi} \frac{1}{\varepsilon_{\mathrm{IR}}} 
\Bigr) \overline{\psi} \fms{p} \psi 
+ \Bigl( 1-\frac{\alpha_s C_F}{4\pi}
\frac{4}{\varepsilon_{\mathrm{IR}}} \Bigr) \bigl(- m\overline{\psi}
\psi \bigr).  
\label{full}
\end{equation}
To match this result onto $\rm{SCET_I}$, we convert $\fms{p}$ to 
$i\fmsl{\D}$ and apply Eqs.~(\ref{psi}), (\ref{comb1}), and
(\ref{comb2}) to Eq.~({\ref{full}}). Then we obtain 
\begin{eqnarray} 
\overline{\psi} \Bigl[\fms{p}-m -\sum(\fms{p},m)\Bigr] \psi
&\longrightarrow& 
\Bigl( 1-\frac{\alpha_s C_F}{4\pi} \frac{1}{\varepsilon_{\mathrm{IR}}}  
\Bigr) \mathcal{K}\nonumber \\
\label{full1}
+\Bigl( 1-\frac{\alpha_s C_F}{4\pi}
\frac{4}{\varepsilon_{\mathrm{IR}}} \Bigr) \mathcal{O}_m^{(1)}&+&
\Bigl( 1-\frac{\alpha_s C_F}{4\pi}
\frac{7}{\varepsilon_{\mathrm{IR}}}  \Bigr) \mathcal{O}_m^{(2)}, 
\end{eqnarray}
where the operators $\mathcal{K}$, $\mathcal{O}_m^{(1)}$, and 
$\mathcal{O}_m^{(2)}$ are defined in Eqs.~(\ref{comb1}) and
(\ref{comb2}), and we use the on-shell renormalization scheme in which
the infrared divergences are regulated by the poles in
$\varepsilon_{\mathrm{IR}}$.

In order to examine if the effective theory reproduces the infrared
divergences of the full theory and to extract the Wilson coefficients
of $\mathcal{O}_m^{(1)}$ and $\mathcal{O}_m^{(2)}$, we need to
calculate the one-loop corrections of $\mathcal{O}_m^{(1)}$ and
$\mathcal{O}_m^{(2)}$ in $\rm{SCET_I}$. For the
strange collinear quark (in the case of up or down quarks the mass
operators are more suppressed), both mass operators are subleading
because the operators start at order $\lambda$ or $\lambda^2$ since
they are given as
\begin{eqnarray}
\mathcal{O}_m^{(1)}&=&m \overline{\xi}_{n,p'}  
\Bigl[ i \fmsl{D}_c^{\perp}, W \frac{1}{\overline{\mathcal{P}}}
W^{\dagger} \Bigr] \nn \xi_{n,p} + \cdots =
\mathcal{O}(\lambda)+\cdots, \nonumber \\ \label{O}
\mathcal{O}_m^{(2)}&=&-m^2 \overline{\xi}_{n,p'}  
W \frac{1}{\overline{\mathcal{P}}} W^{\dagger} \nn \xi_{n,p} + \cdots 
= \mathcal{O}(\lambda^2)+\cdots.
\end{eqnarray}
Here $\overline{\mathcal{P}} = \overline{n} \cdot \mathcal{P}$ and $W$ 
is the collinear Wilson line, 
\begin{equation}
W (x) = \Bigl[\sum_{\rm{perms}} \exp \Bigl(-g 
\frac{1}{\overline{\mathcal{P}}}
\overline{n} \cdot A_{n,q} (x) \Bigr) \Bigr].
\end{equation}
Since the subleading terms in the right side of Eq.~(\ref{O}) are
connected  to the leading terms by the reparameterization invariance
and the gauge symmetries, it is sufficient to consider the loop
corrections of the leading operators which we will denote as
$O_m^{(1)}$ and $O_m^{(2)}$.  

\begin{figure}[t]
\begin{center}
\epsfig{file=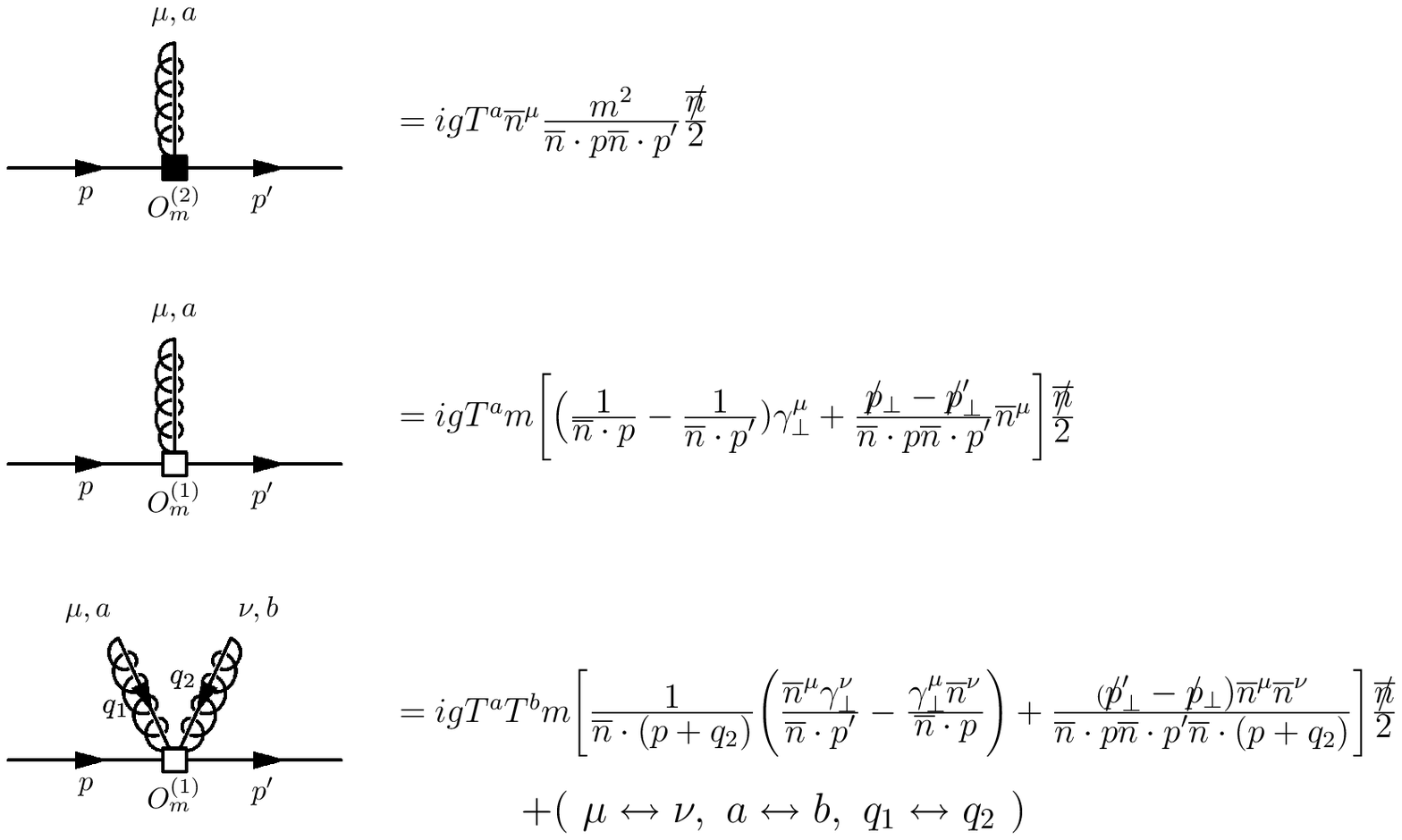, width=13cm}
\end{center}
\vspace{-1.0cm}
\caption{Feynman rules for the operators $O_m^{(1)}$ and $O_m^{(2)}$ with 
one or two collinear gluons.}
\label{fig1}
\end{figure}

\begin{figure}[b]
\begin{center}
\epsfig{file=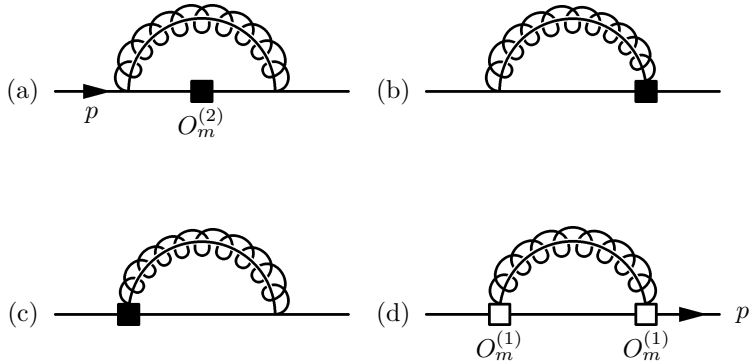, width=10cm}
\end{center}
\vspace{-1.5cm}
\caption{Feynman diagrams for the radiative corrections to $O_m^{(2)}$
  at one loop.}
\label{fig2}
\end{figure}

First let us consider the one loop corrections of $O_m^{(2)}$.
The relevant interaction vertices and their Feynman rules are shown in
Fig.~\ref{fig1}. The Feynman diagrams for the radiative corrections at
one loop are shown in Fig.~\ref{fig2}. The radiative corrections for
$O_m^{(2)}$ with collinear gluons will be the same as those in
Fig.~\ref{fig2} due to the gauge invariance.
If we put the external particle off the mass shell, the radiative
corrections in Fig.~\ref{fig2} are given as
\begin{eqnarray}
M_a^{(2)} &=& -\frac{\alpha_s C_F}{4\pi}  
\frac{m^2}{\overline{n}\cdot p} \nn \Bigl[ -2 -2
\Bigl(\frac{1}{\eps^2} -\frac{1}{\eps} -\frac{\pi^2}{12} \Bigr)
(-p^2)^{-\eps} \Bigr], \nonumber \\
M_b^{(2)} &=& M_c^{(2)} = -\frac{\alpha_s C_F}{4\pi} 
\frac{m^2}{\overline{n} \cdot p} \nn \Bigl(\frac{2}{\eps} +4\Bigr)
(-p^2)^{-\eps},  \nonumber \\
M_d^{(2)} &=&  -\frac{\alpha_s C_F}{4\pi} 
\frac{m^2}{\overline{n} \cdot p} \nn \Bigl( \frac{2}{\eps^2}
+\frac{1}{\eps} +3-\frac{\pi^2}{6} \Bigr)  (-p^2)^{-\eps}. 
\end{eqnarray}
Adding all these, we obtain
\begin{equation}
M_{a}^{(2)}+M_{b}^{(2)}+M_{c}^{(2)}+M_{d}^{(2)}=
-\frac{\alpha_s C_F}{4\pi} \frac{m^2}{\overline{n}\cdot p} \nn \Bigl(
\frac{7}{\eps} +7 \ln \frac{\mu^2}{-p^2} +9\Bigr). 
\label{O2off}
\end{equation}
This result will be used in computing the jet function for
$\overline{B} \rightarrow X_s \gamma$ at order $\alpha_s$. [See Fig.~6
(a), (b) and (c).] 
In pure dimensional regularization, all these diagrams vanish since
the integrals are dimensionless. However, it is transparent to write
the result in pure dimensional regularization from Eq.~(\ref{O2off})
as 
\begin{equation}
M_{a}^{(2)}+M_{b}^{(2)}+M_{c}^{(2)}+M_{d}^{(2)}=
\frac{\alpha_s C_F}{4\pi} \frac{-m^2}{\overline{n}\cdot p} \nn \Bigl(
\frac{7}{\eps} -\frac{7}{\eps_{\mathrm{IR}}}\Bigr),
\label{O2}
\end{equation}
since the ultraviolet divergent piece is known. 

\begin{figure}[b]
\begin{center}
\epsfig{file=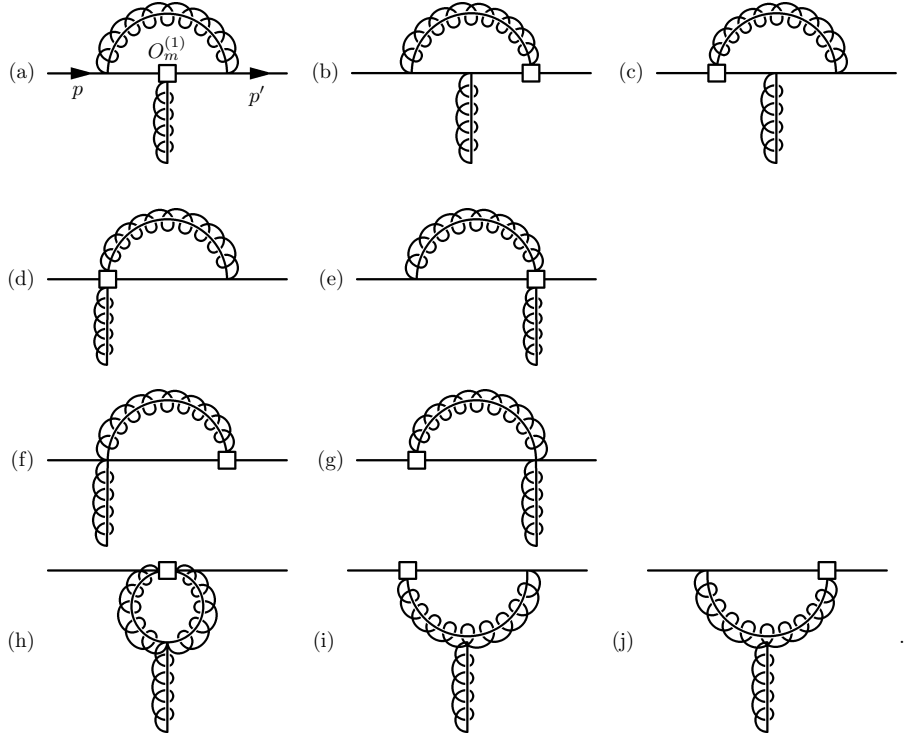, width=12cm}
\end{center}
\vspace{-.5cm}
\caption{Feynman diagrams for one-loop corrections to $O_m^{(1)}$ in
  the background field gauge.}
\label{fig3}
\end{figure}

For the one-loop corrections of the operator $O_m^{(1)}$, 
which has at least one collinear gluon, it is convenient to use the
background gauge field method~\cite{Abbott:1980hw}. 
Since the product of $g$ and the background field $A_n$  
is not renormalized in the background field gauge, 
the number of Feynman diagrams to compute is fairly reduced, and they
are shown in Fig.~\ref{fig3}. The computation of the diagrams is
straightforward using the on-shell dimensional regularization scheme
with the external quark momenta $p^2=p'^2=0$. The results without and
with a triple gluon vertex are given by
\begin{eqnarray}
&&M_{a}^{(1)}+M_{b}^{(1)}+M_{c}^{(1)}+M_{d}^{(1)}
+M_{e}^{(1)}+M_{f}^{(1)}+M_{g}^{(1)} \nonumber \\
&=& 3 \frac{\alpha_s C_F}{4\pi} 
\Bigl(\frac{1}{\eps} - \frac{1}{\eps_{\rm{IR}}}
\Bigr)~m g \overline{\xi}_{n,p'}
\Bigl[\Bigl(\frac{1}{\overline{n}\cdot p}  
-\frac{1}{\overline{n}\cdot p'} \Bigr) \fmsl{A}^{\perp}_n +
\frac{\fms{p}_{\perp}-\fms{p}_{\perp}'}
{\overline{n}\cdot p \overline{n} \cdot p'} \overline{n} \cdot A_n
\Bigr] \nn \xi_{n,p} \nonumber \\ 
&&+ \frac{\alpha_s}{4\pi} \bigl(-\frac{1}{2N}\bigr)
\Bigl(\frac{1}{\eps} - \frac{1}{\eps_{\rm{IR}}}
\Bigr)~m g \overline{\xi}_{n,p'} \Bigl[\bigl(\frac{1}{\overline{n}\cdot p} 
-\frac{1}{\overline{n}\cdot p'} \bigr) \fmsl{A}^{\perp}_n +
\frac{\fms{p}_{\perp}-\fms{p}_{\perp}'}
{\overline{n}\cdot p \overline{n} \cdot p'} \overline{n} \cdot A_n \Bigr] \nn
\xi_{n,p}, \\ 
&&M_{h}^{(1)}+M_{i}^{(1)}+M_{j}^{(1)} \nonumber \\
&=& \frac{\alpha_s}{4\pi} \frac{N}{2} 
\Bigl(\frac{1}{\eps} - \frac{1}{\eps_{\rm{IR}}}
\Bigr)mg \overline{\xi}_{n,p'}
\Bigl[\Bigl(\frac{1}{\overline{n}\cdot p} 
-\frac{1}{\overline{n}\cdot p'} \Bigr) \fmsl{A}^{\perp}_n +
\frac{\fms{p}_{\perp}-\fms{p}_{\perp}'}
{\overline{n}\cdot p \overline{n} \cdot p'} \overline{n} \cdot A_n \Bigr] \nn
\xi_{n,p}, 
\end{eqnarray} 
where $M_i^{(1)}$ represents the $i$th diagram in Fig.~\ref{fig3}, and
$N$ is the number of colors. Summing these two results  with $C_F =
(N^2-1)/(2N)$, the radiative correction of the operator $O_m^{(1)}$ at
one loop is given as
\begin{eqnarray}
M^{(1)}&=&\frac{\alpha_s C_F}{4\pi} \Bigl(\frac{4}{\eps} -
\frac{4}{\eps_{\rm{IR}}} \Bigr)~m g \overline{\xi}_{n,p'}
\Bigl[\bigl(\frac{1}{\overline{n}\cdot p} -\frac{1}{\overline{n}\cdot
  p'} \bigr) \fmsl{A}^{\perp}_n +
\frac{\fms{p}_{\perp}-\fms{p}_{\perp}'} {\overline{n}\cdot p
  \overline{n} \cdot p'} \overline{n} \cdot A_n \Bigr] \nn \xi_{n,p}
\nonumber \\  \label{O1}
&=& \frac{\alpha_s C_F}{4\pi} 
\Bigl(\frac{4}{\eps} - \frac{4}{\eps_{\rm{IR}}}\Bigr) O_m^{(1)}.
\end{eqnarray}

From Eqs.~(\ref{O2}) and (\ref{O1}), we can see that the radiative
corrections of the operators $O_m^{(2)}$ and $O_m^{(1)}$ in
$\mathrm{SCET}_{\mathrm{I}}$ reproduce the infrared divergences in the
full theory. And since the radiative corrections are the same in both
theories, the Wilson coefficients of both operators are 1 with no
contribution at one loop. We can also extract the counterterm for the
quark mass in $\mathrm{SCET}_{\mathrm{I}}$. The counterterm $Z_{\xi}$
for the wavefunction renormalization of a collinear quark is given by   
\begin{equation}
Z_{\xi} = 1 - \frac{\alpha_s C_F}{4\pi} \frac{1}{\eps},
\end{equation}
which is the same as the counterterm in the full theory for the quark
field. Therefore we obtain the counterterm for the quark mass from
$O_m^{(1)}$ and $O_m^{(2)}$ as
\begin{equation}
Z_m^{\rm{SCET_I}} = 1- \frac{\alpha_s C_F}{4\pi}\frac{3}{\eps},
\label{mass}
\end{equation}
which is the same as the full theory mass renormalization to
first order in $\alpha_s$. This is to be expected since we do not 
integrate out any degrees of freedom relevant to the collinear quark
mass from the matching. In summary, it has been shown that the
counterterms for the wavefunction and the quark mass are the same as
those in the full theory, and there are no contributions to the
coefficients of the operators at one loop; that is, the operators are
not renormalized to order $\alpha_s$. 

The matching between $\rm{SCET_I}$ and $\rm{SCET_{II}}$ is trivial because 
there is no hard-collinear degrees of freedom 
($p_{hc}^2 \sim Q \Lambda$) to be integrated out in the
$\rm{SCET_I}$ Lagrangian. Note that the situation is different for
heavy-to-light currents with the spectator interactions in $B$ decays
and for soft-collinear currents \cite{Becher:2003kh,Chay:2004va},
in which  there arise nontrivial Wilson coefficients (or jet
functions) from the matching between $\rm{SCET_I}$ and
$\rm{SCET_{II}}$. A more concrete analysis on the hard-collinear modes
is discussed in Refs.~\cite{Chay:2004va,Beneke:2003pa}.
However the operators $O_m^{(1)}$ and $O_m^{(2)}$ in
$\rm{SCET_{II}}$ remain as they are in $\rm{SCET_I}$  since the
collinear momentum $p_c^2 = m^2$ is still very small compared to
matching scale  $\mu \sim \sqrt{Q\Lambda}$. Therefore in
$\mathrm{SCET}_{\mathrm{II}}$, the mass operators are regarded as the
leading operators for the strange collinear quark, and   
the operators have the same Wilson coefficients and
the same renormalization behavior as in $\rm{SCET_I}$ with the same 
mass renormalization given by Eq.~(\ref{mass}).

\section{Quark mass corrections to $\bar{B} \to X_s \gamma$ decays}
Inclusive $B$ decays based on HQET \cite{InHQET} have been widely
studied to extract Cabibbo-Kobayashi-Maskawa (CKM) matrix elements and
to search for possible new physics. When an emitted photon is
energetic in the region of the phase space with $p_X^2 \sim m_B \Lambda$,
SCET along with HQET is applicable and has been successfully
applied~\cite{Bauer:2000ew,Inclusive}. In this case, the differential
decay rate can be given by a factorized form as
\begin{equation} 
\frac{d \Gamma}{d E_{\gamma}} \propto H ~ J \otimes f,
\label{facto}
\end{equation}
where $\otimes$ means the appropriate convolution. Here $H$ is a hard
factor obtained from the matching between the full theory and
$\rm{SCET_I}$, $J$ is a jet function obtained by integrating out  
hard-collinear objects, and $f$ represents the shape function of a $B$
meson, which consists of only soft interactions and is purely
nonperturbative. 
 
Recently the corrections of order $\Lambda/m_b$ to $\bar{B}\to
X_s \gamma$ and $\bar{B} \to  X_u l \overline{\nu}~$ decays in the
endpoint region have been investigated using SCET 
\cite{Lee:2004ja,Bosch:2004cb,Beneke:2004in}.  
Here the factorization formula Eq.~(\ref{facto}) still holds, and the 
subleading shape functions are studied to clarify the uncertainty from the 
theoretical analysis. When the effect of the strange quark mass 
of order $\Lambda$ is included in $\bar{B} \to X_s \gamma$ or $\bar{B}
\to X_s l \overline{l}$, the mass corrections can also give a
nonnegligible contribution of order $\Lambda/m_b$. In this section we
focus on this fact and analyze the mass corrections to the decay
$\bar{B} \to X_s \gamma$ in the endpoint region. The result is also
applicable to the $\bar{B} \to X_s l \overline{l}$, but it 
is not considered here.   

The effective weak Hamiltonian for $\bar{B} \to X_s \gamma$ is given
by  \cite{Grinstein:1990tj}
\begin{equation} 
\mathcal{H}_{\rm{eff}} = - 4 \frac{G_F}{\sqrt{2}} V_{tb}V_{ts}^* 
\sum_{i=1}^{8} C_i^{\rm{full}}(\mu) Q_i (\mu),  
\end{equation}   
where the main contribution comes from the operator
\begin{equation} 
Q_7 = \frac{e}{16\pi^2} \overline{s} \sigma_{\mu \nu} F^{\mu \nu} 
\bigl( m_b P_R + m_s P_L \bigr) b.    
\end{equation} 
Here $P_{R,L} = (1 \pm \gamma_5)/2$ and $F^{\mu \nu}$ is the
electromagnetic field strength tensor. We choose the frame in which  
the photon momentum $q^{\mu}$ is in the $\overline{n}^{\mu}$
direction, $q^{\mu} = n\cdot q \overline{n}^{\mu}/2 = E_{\gamma}
\overline{n}^{\mu}$, where the photon energy 
$E_{\gamma}$ near the endpoint satisfies $m_B - 2 E_{\gamma}
\lesssim \Lambda$. The strange quark can be taken as a collinear quark
in the $n^{\mu}$ direction in the rest frame of a $B$ meson.  

Let us define the forward scattering amplitude $T_{\mu\nu}$ as
\begin{equation}
T_{\mu\nu} = \frac{1}{2m_B} \langle \overline{B} |\hat{T}_{\mu \nu}
|\overline{B} \rangle,  
\end{equation}
where $\hat{T}_{\mu\nu}$ is given by
\begin{equation}
\hat{T}_{\mu\nu} = -i \int d^4 z e^{-i q\cdot z} T[ J_{\mu}^{\dagger}
  (z) J_{\nu} (0)]  
\end{equation}
with the current
\begin{equation} 
J^{\mu} = i~ \overline{s} \sigma_{\mu \nu} q^{\nu} P_R b  +
\frac{m_s}{m_b}~ i ~\overline{s} \sigma_{\mu \nu} q^{\nu} P_L b. 
\label{J}
\end{equation}  
The inclusive photon energy spectrum can be written as  
\begin{equation} 
\frac{1}{\Gamma_0} \frac{d \Gamma}{d E_{\gamma}} 
= \frac{8 E_{\gamma}}{m_B^3} \frac{1}{\pi} 
\mathrm{Im} T_{\mu}^{\mu}(E_{\gamma}), 
\label{rate}
\end{equation}
where
\begin{equation} 
\Gamma_0 = \frac{G_F^2 m_B^3 m_b^2}{32 \pi^4} \alpha |V_{tb} V_{ts}^*|^2 
|C_7^{\mathrm{full}}(m_b)|^2.
\end{equation}
The forward scattering amplitude $T_{\mu\nu} (E_{\gamma})$
in SCET near the endpoint region is given by the factorized form
in Eq.~(\ref{facto}) and the power counting can be performed
systematically. The hard part can be computed from the matching
between the full theory and $\rm{SCET_I}$, and the heavy-to-light
current can be expanded in terms of the currents in
$\mathrm{SCET}_{\mathrm{I}}$ in powers of $\lambda \sim
\sqrt{\Lambda/m_b}$. Then the time-ordered product of the effective
currents can be expressed as a convolution of the jet function and the
shape function of the $B$ meson by matching onto $\rm{SCET_{II}}$.
As a result, the forward scattering amplitude is given by the
convolution of the hard part, the jet function, and the shape
functions. 

We investigate the strange quark mass corrections 
to the inclusive decay rate to first order in $\Lambda/m_b$ and
$\alpha_s$. We show that these corrections can also be written in a
factorized form and the mass corrections reside only in the jet
functions. This mass correction should be included in the subleading
contribution along with other subleading corrections from the shape 
function to order $\Lambda/m_b$, which was extensively discussed in 
Refs.~\cite{Lee:2004ja,Bosch:2004cb,Beneke:2004in}.  
 
\subsection{Matching a heavy-to-light current with a quark mass} 
Let us consider matching the heavy-to-light tensor current 
$J_{\mu \nu} = \overline{s} \sigma_{\mu \nu} (1+\gamma_5) b$ 
at $\mu \sim m_b \sim \overline{n}\cdot p$ where 
$\overline{n}\cdot p$ is the large momentum component of the collinear strange quark. 
The full-theory current can be matched onto the currents in
$\rm{SCET_I}$, in which the hard degrees of freedom such as $m_b$ and the
large off-shellness $p_{\mathrm{hard}}^2 \sim m_b^2 \sim m_b
\overline{n}\cdot p$ are integrated out. By choosing the heavy
quark velocity as $v_{\perp} = 0, ~n \cdot v = \overline{n} \cdot v =1$, 
the heavy-to-light current can be expanded in $\rm{SCET_I}$ as 
\begin{eqnarray} 
J_{\mu \nu} = e^{i (\tilde{p} \cdot z - m_b v\cdot z)}
&\Bigl\{& \sum_{i} \int d\omega C_i (\omega) j_{i \mu \nu}^{(0)}(\omega)
       + \sum_{i} \int d\omega B_i (\omega) j_{i \mu \nu}^{(1)}(\omega)  
\nonumber \\ 
\label{current}  
&&+\sum_{i} \int d\omega A_i' (\omega) j_{i \mu \nu}^{(2)}(\omega)
+\sum_{i} \int d\omega A_i (\omega) j_{i \mu \nu}^{(m)}(\omega) + \cdots ~~
\Bigr\},
\end{eqnarray}
where the superscripts $k$ ($k=0,1,2$) denote the order in $\lambda$, and 
another superscript $m$ indicates the operators with the strange quark
mass. The currents $j_{i \mu \nu}^{(m)}$ are of the same order as 
$j_{i \mu \nu}^{(2)}$ as long as the mass is regarded as $m \sim \Lambda$.
From now on, we  suppress the exponential factors with the
understanding that the label momenta are conserved. 
Since we focus on the mass corrections of the heavy-to-light 
currents and their relations to the leading or the subleading currents
in $\lambda$, we will not consider the currents $j_{i \mu \nu}^{(2)}$ 
any more.  The detailed analysis on these currents can be found
in Ref.~\cite{Lee:2004ja}.

At tree level, the current operator in the full theory can be
expressed in terms of the currents in $\mathrm{SCET}_{\mathrm{I}}$ as   
\begin{eqnarray} \label{tree}
J_{\mu \nu} &=& \overline{s} \sigma_{\mu \nu} (1+\gamma_5) b 
\nonumber \\ &\longrightarrow& 
\overline{\xi}_n W \s (1+ \gamma_5) h_v
+ \overline{\xi}_n \nn i \overleftarrow{\fmsl{D}^{\perp}_c} W 
\frac{1}{\overline{n} \cdot \mathcal{P}^{\dagger}}  \s (1+ \gamma_5) h_v
\nonumber \\ 
&&+ \frac{1}{m_b} \overline{\xi}_n \s (1+\gamma_5) i \fmsl{D}_c^{\perp} 
W \frac{\fms{n}}{2} h_v + m \overline{\xi}_n W \nn
\frac{1}{\overline{n} \cdot \mathcal{P}^{\dagger}} \s 
 (1+\gamma_5) h_v \\
&=& j_{1\mu\nu}^{(0)} +  j_{1\mu\nu}^{(1)} + j_{2\mu\nu}^{(1)}
    + j_{1\mu\nu}^{(m)}, \nonumber \\
\tilde{J}_{\mu\nu} &=& \frac{m}{m_b} \overline{s} \sigma_{\mu\nu}
    (1-\gamma_5) b \longrightarrow \frac{m}{m_b} \overline{\xi}_n
    \sigma_{\mu\nu} (1-\gamma_5) h_v = j_{3\mu\nu}^{(m)}, 
\end{eqnarray}
where $\xi_n$ is a collinear strange quark field and $h_v$ is a heavy 
quark field. 

At order $\alpha_s$, we employ the modified 
minimal subtraction ($\overline{\mathrm{MS}}$) scheme using 
on-shell dimensional regularization. In the full theory, 
the matrix element of the tensor current $J_{\mu \nu}$ at one loop is
given as  
\begin{eqnarray}\label{full2}
\langle J_{\mu \nu} 
\rangle^{(1)}&=&\frac{\alpha_s C_F}{4\pi}
\Bigl\{\Bigl[ -\frac{1}{\eps^2} - \frac{2}{\eps} + \frac{2}{\eps} 
\ln\frac{\overline{n}\cdot p}{\mu} - 4 \ln \frac{\mu}{m_b} + 
\frac{2(1-2x)}{1-x} \ln x \nonumber  \\ 
&-& 2\ln^2 \frac{\overline{n}\cdot p}{\mu} - \frac{\pi^2}{12} - 2 Li_2(1-x)
  -4 \Bigr]  \langle \overline{s} \s (1+\gamma_5) b \rangle \nonumber \\ 
&+&\Bigl[\frac{4}{1-x} \ln x \Bigr]~\frac{1}{m_b} 
\langle i\overline{s} 
(\gamma_{\mu} p_{\nu} - \gamma_{\nu} p_{\mu}) (1- \gamma_5) b \rangle
\nonumber \\ 
&+&\Bigl[ \frac{2}{x}\bigl( \frac{1}{\eps} 
 -2 \ln \frac{\overline{n} \cdot p}{\mu}+2\bigr)\Bigl] \frac{m}{m_b} 
 \langle \overline{s} \s (1-\gamma_5) b \rangle \nonumber \\ 
&-&\Bigl[\frac{4}{x}\bigl(\frac{1}{\eps}
-2 \ln \frac{\overline{n} \cdot p}{\mu}+2-\frac{x}{1-x} \ln x \bigr)\Bigl] 
\frac{m}{m_b^2} \langle i\overline{s} 
(\gamma_{\mu} p_{b\nu} - \gamma_{\nu} p_{b\mu}) (1+\gamma_5) b \rangle  
\nonumber \\ 
&-&\frac{2m}{xm_b^2}
\langle i\overline{s} 
(\gamma_{\mu} p_{\nu} - \gamma_{\nu} p_{\mu}) (1+ \gamma_5) b 
\rangle\Bigr\}, 
\end{eqnarray}
where $x=\overline{n}\cdot p/m_b$, and all the poles in $1/\eps$
represent the IR divergences. Here we use the equations of motion
$\fms{p}_b b = m_b b$ and $\overline{s} \fms{p}  = m \overline{s}$
putting each quark on shell with $p_b^2 = m_b^2$, $p^2 = m^2 \to 0$,
keeping the terms to first order in the strange quark mass $m$. 
For $\tilde{J}_{\mu\nu}$ the matrix element at one loop is given
by 
\begin{eqnarray}
\langle \tilde{J}_{\mu\nu} \rangle^{(1)} &=& \frac{\alpha_s C_F}{4
  \pi}  \Bigl\{\Bigl[ -\frac{1}{\eps^2} - \frac{2}{\eps} + \frac{2}{\eps} 
\ln\frac{\overline{n}\cdot p}{\mu} - 4 \ln \frac{\mu}{m_b} + 
\frac{2(1-2x)}{1-x} \ln x \nonumber \\ 
&-& 2\ln^2 \frac{\overline{n}\cdot p}{\mu} - \frac{\pi^2}{12} - 2 Li_2(1-x)
  -4 \Bigr] \frac{m}{m_b} \langle \overline{s} \s (1-\gamma_5) b
  \rangle \nonumber \\  
&+& \frac{4m}{m_b^2} \frac{\ln x}{1-x}\langle i\overline{s}
  (\gamma_{\mu} p_{\nu}   -\gamma_{\nu} p_{\mu}) (1+\gamma_5)
  b\rangle  \Bigr\}.
\end{eqnarray}

Now we expand the current operators in Eq.~(\ref{full2}) in powers of
$\lambda$ using the momentum decomposition 
$p^{\mu}=\overline{n}\cdot p n^{\mu} /2 + p_{\perp}^{\mu} +n\cdot p
\overline{n}^{\mu}/2$. These operators can be written in terms of the
gauge-invariant effective currents as 
\begin{eqnarray} 
&&\overline{s} \sigma_{\mu \nu} (1+\gamma_5) b 
\longrightarrow  C_1 \overline{\xi}_n W \s (1+ \gamma_5) h_v
+B_1 \overline{\xi}_n \nn i \overleftarrow{\fmsl{D}^{\perp}_c} W 
\frac{1}{\overline{n} \cdot \mathcal{P}^{\dagger}}  \s (1+ \gamma_5) h_v
\nonumber \\ 
&& \hspace{2.0cm} +B_2 \frac{1}{m_b} \overline{\xi}_n \s (1+\gamma_5) i
\fmsl{D}_c^{\perp} W \frac{\fms{n}}{2} h_v + A_1 m \overline{\xi}_n W \nn
\frac{1}{\overline{n} \cdot \mathcal{P}^{\dagger}} \s 
 (1+\gamma_5) h_v + \cdots \nonumber \\
&& \hspace{2.0cm} =  C_1j_{1\mu\nu}^{(0)} + B_1 j_{1\mu\nu}^{(1)} +
B_2 j_{2\mu\nu}^{(1)} 
    + A_1  j_{1\mu\nu}^{(m)} + \cdots, \nonumber \\ 
&&\frac{2}{x m_b} i\overline{s} (\gamma_{\mu} p_{\nu} - \gamma_{\nu}
p_{\mu}) (1- \gamma_5) b  \longrightarrow  
C_2 i \overline{\xi}_n W (\gamma_{\mu} n_{\nu} - \gamma_{\nu} n_{\mu})
(1-\gamma_5) h_v \nonumber \\
&& \hspace{2.0cm} + B_3 i \overline{\xi}_n \nn i
\overleftarrow{\fmsl{D}^{\perp}_c} W   
\frac{1}{\overline{n} \cdot \mathcal{P}^{\dagger}}  
(\gamma_{\mu} n_{\nu} - \gamma_{\nu} n_{\mu}) (1- \gamma_5) h_v
\nonumber \\
&&\hspace{2.0cm} +2B_4 i\overline{\xi}_n \bigl(\gamma_{\mu} i
\overleftarrow{D_c}^{\perp}_{\nu} - i\overleftarrow{D_c}^{\perp}_{\mu}
\gamma_{\nu} \bigr) 
\frac{1}{\overline{n} \cdot \mathcal{P}^{\dagger}} (1-\gamma_5) h_v 
\nonumber \\
&&\hspace{2.0cm} +A_2i m\overline{\xi}_n \nn  W 
\frac{1}{\overline{n} \cdot \mathcal{P}^{\dagger}}  
(\gamma_{\mu} n_{\nu} - \gamma_{\nu} n_{\mu}) (1- \gamma_5) h_v 
+ \cdots \nonumber \\
\label{expa}
&&\hspace{2.0cm} = C_2 j_{2\mu\nu}^{(0)} + B_3 j_{3\mu\nu}^{(1)} + 2B_4
j_{4\mu\nu}^{(1)} + A_2 j_{2\mu\nu}^{(m)} + \cdots, \\
&&\frac{m}{m_b} \overline{s} \s (1-\gamma_5) b \longrightarrow  
(A_3+\tilde{A}_3) \frac{m}{m_b} \overline{\xi}_n W \s (1-\gamma_5) h_v
+ \cdots = (A_3 +\tilde{A}_3)  j_{3\mu\nu}^{(m)} + \cdots, \nonumber \\  
&&\frac{m}{m_b^2} i\overline{s} 
(\gamma_{\mu} p_{b\nu} - \gamma_{\nu} p_{b\mu}) (1+ \gamma_5) b 
\longrightarrow A_4 \frac{m}{m_b}
i \overline{\xi}_n W (\gamma_{\mu} v_{\nu} - \gamma_{\nu} v_{\mu})
(1+\gamma_5) h_v = A_4 j_{4\mu\nu}^{(m)} + \cdots, \nonumber \\  
&&\frac{2m}{x m_b^2}
i\overline{s} (\gamma_{\mu} p_{\nu} - \gamma_{\nu} p_{\mu}) (1+ \gamma_5) b  
\longrightarrow (A_5 +\tilde{A}_5)  \frac{m}{m_b}
i \overline{\xi}_n W (\gamma_{\mu} n_{\nu} - \gamma_{\nu} n_{\mu})
(1+\gamma_5) h_v \nonumber \\
&&\hspace{2.0cm} =( A_5 +\tilde{A}_5)  j_{5\mu\nu}^{(m)} + \cdots,
\nonumber  
\end{eqnarray}
where we keep the effective currents to $\mathcal{O}(\lambda^2)$. Here
we use the fact that 
\begin{equation} 
\tilde{J}_{\mu \nu} = \frac{m}{m_b} \overline{s} \sigma_{\mu \nu}
(1-\gamma_5) b  \longrightarrow \tilde{A_3}~ j^{(m)}_{3\mu\nu} 
+\tilde{A_5}~ j^{(m)}_{5\mu\nu}. 
\label{subc}
\end{equation}

All the Wilson coefficients at tree level are 0 except 
\begin{equation}
C_1 = B_1 = A_1=B_2 = \tilde{A}_3 =1.  
\end{equation}
Due to the reparameterization invariance, some of the Wilson
coefficients at subleading order are related to the leading
coefficients. They are given by
\begin{equation}
C_1 = B_1 = A_1, \ C_2 = B_3 =2B_4 =A_2
\end{equation}
to all orders in $\alpha_s$ since those subleading operators with the
corresponding Wilson coefficients are obtained by expanding the
collinear field in a reparameterization-invariant way. However the
subleading operator with $B_2$ is an
independent operator. This operator can be obtained when we consider
the heavy-to-light current in which a collinear gluon is emitted from
the heavy quark and we integrate out the intermediate-state heavy
particle. Therefore the coefficient $B_2$ is not related to other
Wilson coefficients and should be computed independently. The
radiative correction for the operator with $B_2$ was considered at one loop in
Refs.~\cite{Beneke:2004rc,Hill:2004if}. And $\tilde{A}_3$ and
$\tilde{A}_5$ come from the operator proportional to $m_s=m$ in
$Q_7$, while $A_3$ and $A_5$ come
from the subleading contribution of the leading operator in
$Q_7$ at higher orders in $\alpha_s$.   

\begin{figure}[t]
\begin{center}
\epsfig{file=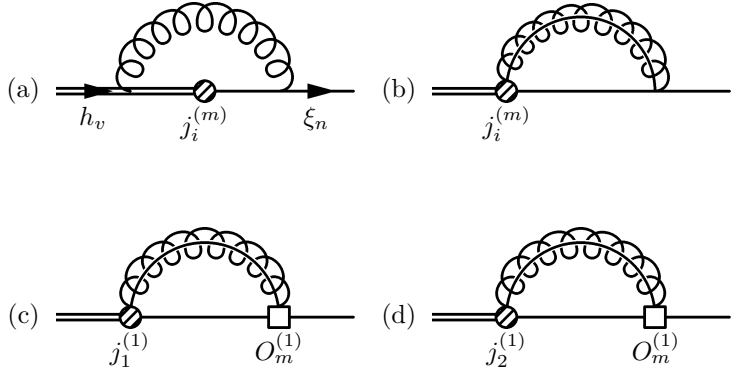, width=10cm}
\end{center}
\vspace{-1.5cm}
\caption{Feynman diagrams for the mass correction to the
  heavy-to-light current with   $j_i^{(m)}$ ($i=1,2,\cdots,5$) and
  $O_m^{(1)}$.}
\label{fig4}
\end{figure}

In order to match the full theory onto $\mathrm{SCET}_{\mathrm{I}}$,
we compute the radiative corrections in
$\mathrm{SCET}_{\mathrm{I}}$. The relevant Feynman diagrams in
$\mathrm{SCET}_{\mathrm{I}}$ at one loop are  
shown in Fig.~\ref{fig4}. We have verified that the infrared 
divergences of the full theory in Eq.~(\ref{full2}) are fully
reproduced in the effective theory from the explicit calculations of the
diagrams with the self energy of the external quarks, and they cancel
out in matching. In computing the Wilson coefficients, the point is
that all the radiative corrections in the effective theory are simply
zero when the on-shell dimensional regularization scheme is employed,
and the Wilson coefficients can be easily obtained.

 The difference of the
residues in the wave function renormalization between the full theory
and the effective theory for the heavy quark  at one loop is given by 
\begin{equation} 
\frac{1}{2}\Bigl(R_b^{(1)}-R_h^{(1)}\Bigr) 
= - \frac{\alpha_s C_F}{4\pi}\Bigl( 3 \ln \frac{\mu}{m_b} + 2 \Bigr), 
\end{equation}    
and we find the Wilson coefficients $C_i$ for $j_{i\mu\nu}^{(0)}$,
$B_i$ for $j_{i\mu\nu}^{(1)}$, and $A_i$, $\tilde{A}_i$ for
$j_{i\mu\nu}^{(m)}$ as  
\begin{eqnarray} \label{wil} 
C_1=B_1=A_1&=& 1+ \frac{\alpha_s C_F}{4\pi} 
\Bigl[-6 - 7 \ln \frac{\mu}{m_b} + \frac{2(1-2x)}{1-x} \ln x 
 - 2 \ln^2 \frac{\overline{n}\cdot p}{\mu} \nonumber \\ 
&-& 2 Li_2 (1-x) - \frac{\pi^2}{12}
\Bigr], \nonumber \\ 
C_2=B_3=2B_4=A_2&=& \frac{\alpha_s C_F}{4\pi} \Bigl( \frac{2x}{1-x} \ln 
x\Bigr), \nonumber \\
A_3&=& \frac{\alpha_s C_F}{4\pi} \Bigl( - \frac{4}{x} 
\ln\frac{\overline{n}\cdot p}{\mu} + \frac{4}{x} \Bigr),  \nonumber \\ 
A_4&=& \frac{\alpha_s C_F}{4\pi} \Bigl(  \frac{8}{x} 
\ln\frac{\overline{n}\cdot p}{\mu} - \frac{8}{x} 
+ \frac{4}{1-x} \ln x \Bigr),~~~A_5= -
\frac{\alpha_s C_F}{4\pi}, \nonumber  \\
\tilde{A_3} &=& 1+ \frac{\alpha_s C_F}{4\pi} 
\Bigl[-6 - 7 \ln \frac{\mu}{m_b} + \frac{2(1-2x)}{1-x} \ln x 
 - 2 \ln^2 \frac{\overline{n}\cdot p}{\mu} \nonumber \\
&-& 2 Li_2 (1-x) - \frac{\pi^2}{12} \Bigr], \nonumber \\ 
\tilde{A_5} &=&  
\frac{\alpha_s C_F}{4\pi} \Bigl( \frac{2x}{1-x} \ln x\Bigr).
\end{eqnarray} 
The Wilson coefficients $C_1(\mu)$ and $C_2(\mu)$ are basically
identical to those obtained in Ref.~\cite{Bauer:2000yr} although the
operator basis is different. The Wilson coefficients $A_i$,
$\tilde{A}_i$ and $B_i$ except $B_2$ are new and first calculated
here.    

Note that all the operators in the basis
$\{j_{1\mu\nu}^{(m)},\cdots,j_{5\mu\nu}^{(m)}\}$ are not independent. 
Because $\nn =\fms{v} - \frac{\fms{n}}{2}$ in the $B$ meson rest
 frame with the choice of $v_{\perp}^{\mu} = 0$, the operator 
$j_{1\mu\nu}^{(m)}$ can be written as  
\begin{eqnarray} 
j_{1\mu\nu}^{(m)}&=&m~\overline{\xi}_n W (\fms{v} - \frac{\fms{n}}{2})
\frac{1}{\overline{n} \cdot \mathcal{P}^{\dagger}} \s (1+\gamma_5) h_v 
\nonumber \\  
&=& m \overline{\xi}_n W 
\frac{1}{\overline{n} \cdot \mathcal{P}^{\dagger}} \s (1-\gamma_5) h_v 
-2m i \overline{\xi}_n W (\gamma_{\mu} v_{\nu} - \gamma_{\nu} v_{\mu})
\frac{1}{\overline{n} \cdot \mathcal{P}^{\dagger}} \s (1+\gamma_5) h_v \\
&=& \frac{1}{x} \Bigl( j_{3\mu\nu}^{(m)} - 2 j_{4\mu\nu}^{(m)}\Bigr).
\nonumber
\end{eqnarray}
Therefore the number of the independent operators in the basis 
is four. But it is useful to use this basis because the
reparameterization invariance is shown transparently, 
as shown in Eq.~(\ref{wil}).

\subsection{Jet functions and factorization in $\rm{SCET_{II}}$}
Let us consider the contribution of the quark mass  
to the forward scattering amplitude $T_{\mu\nu}
(E_{\gamma})$. The current $J_{\mu}$ in Eq.~(\ref{J}) can be written  
in the effective theory as 
\begin{eqnarray} 
J_{\mu} &=& iE_{\gamma} \frac{\overline{n}^{\nu}}{2} J_{\mu\nu} \nonumber \\
\label{j} 
&=& i E_{\gamma} e^{i (\tilde{p} \cdot z - m_b v\cdot z)}
\Bigl\{ \sum_{i} \int d\omega C_i (\omega) j_{i \mu}^{(0)}(\omega)
       + \sum_{i} \int d\omega B_i (\omega) j_{i \mu }^{(1)}(\omega)  
\\
&&+\sum_{i} \int d\omega A_i (\omega) j_{i \mu }^{(m)}(\omega) + \cdots ~~
\Bigr\}, \nonumber 
\end{eqnarray} 
where $j_{i \mu}^{(j)}(\omega) = \frac{\overline{n}^{\nu}}{2} 
j_{i\mu\nu}^{(j)}(\omega)$, ($j=0,1,m$). Here we express
$j_{i\mu\nu}^{(j)}(\omega)$ as  
\begin{equation} 
j_{i\mu\nu}^{(j)}(\omega) = \overline{\xi}_n W 
\delta(\omega - \overline{n}\cdot \mathcal{P}^{\dagger})
\Gamma_i^{(j)} h_v, 
\label{jmunu}
\end{equation} 
with a delta function. In general, the operators
$j_{2\mu\nu}^{(1)}(\omega)$ need  additional parameters
$\omega^{\prime}$ at higher orders in $\alpha_s$ 
since it consists of at least three external particles  
including a collinear gluon, but it is not necessary at one loop since
it is sufficient to consider the tree-level Wilson coefficients of
$j_{2\mu\nu}^{(1)}(\omega)$.  

The forward scattering amplitude  $T_{\mu\nu}(E_\gamma)$ can be
written as 
\begin{equation} 
T_{\mu\nu}(E_{\gamma}) =  \frac{E_{\gamma}^2}{2} \langle
\overline{B}_v | \hat{T}_{\mu\nu} | \overline{B}_v \rangle,
\label{teff}
\end{equation} 
with the normalization of the $B$ meson states in HQET. 
$\hat{T}_{\mu\nu}$ is given by  
\begin{equation} 
\hat{T}_{\mu\nu} = -i \sum_{i,i',k,k'} \int d\omega d\omega' 
C_{i'}^{(k')}(\omega') C_i^{(k)}(\omega) \int d^4 z e^{-ir\cdot z} 
T \Bigl[j_{i'\mu}^{(k')\dagger}(\omega',z),
j_{i\nu}^{(k)}(\omega,0)\Bigr] + \cdots,
\label{tmunu} 
\end{equation}
where $C_i^{(k)}$ ($k=0,1,m,\cdots$) are the Wilson coefficients
$C_i$, $B_i$, $A_i$, and $\tilde{A}_i$ in Eq.~(\ref{j}). The momentum
$r$ in the exponential factor is defined as 
\begin{equation} 
r^{\mu} = q^{\mu} +\tilde{p}^{\mu} -m_b v^{\mu}. 
\end{equation} 
Since the photon momentum $q^{\mu}$ is given by
$q^{\mu}=n \cdot q \overline{n}^{\mu}/2$, the label momentum of the 
collinear strange quark is fixed as 
\begin{equation} 
\overline{n} \cdot p = m_b, ~~~ p_{\perp}^{\mu} = 0   
\end{equation}
giving $\overline{n}\cdot r = r_{\perp} = 0$.  $n\cdot r$ can
be written as 
\begin{equation} 
n\cdot r = n\cdot q - m_b = m_B-n\cdot p_X - m_b,
\label{nr}
\end{equation}  
where $p_X$ is a momentum of the jet $X$ and we use the momentum conservation 
$m_B v^{\mu} = q^{\mu} + p_X^{\mu}$. Evidently $n\cdot r$ is of order
$\Lambda$ since the mass difference $\overline{\Lambda}=m_B-m_b$ and
$n\cdot p_X$ are of order $\Lambda$. 

We can express Eq.~(\ref{tmunu}) showing the dependence of the quark
mass explicitly as 
\begin{eqnarray}  \label{tm1}
\hat{T}_{\mu\nu}^{(m)} &=&- i \sum_{i,i'} \Biggl\{\int d\omega d\omega' 
C_{i'}(\omega') A_i(\omega) \int d^4 z e^{-ir\cdot z} 
T\Bigl[ j_{i'\mu}^{(0)\dagger}(\omega',z),
j_{i\nu}^{(m)}(\omega,0)\Bigr]  \nonumber \\
&+&\int d\omega d\omega' 
C_{i'}(\omega') B_i(\omega) \int d^4 z d^4 x  e^{-ir\cdot z} 
T\Bigl[ j_{i'\mu}^{(0)\dagger}(\omega',z), i\mathcal{L}_m^{(1)}(x), 
 j_{i\nu}^{(1)}(\omega,0) \Bigr]  \nonumber \\
&+&\int d\omega d\omega' 
C_{i'}(\omega') C_i(\omega) \Bigl\{ \int d^4 z d^4 x  e^{-ir\cdot z} 
T\Bigl[j_{i'\mu}^{(0)\dagger}(\omega',z), i\mathcal{L}_m^{(2)}(x), 
j_{i\nu}^{(0)}(\omega,0) \Bigr] \nonumber     \\
&+ &\int d^4 z d^4 x d^4 y e^{-ir\cdot z} 
T\Bigl[ j_{i'\mu}^{(0)\dagger}(\omega',z), i\mathcal{L}_m^{(1)}(x), 
i\mathcal{L}_m^{(1)}(y), j_{i\nu}^{(0)}(\omega,0) \Bigr]
\Bigr\}+\rm{h.c.} \Biggr\}, 
\end{eqnarray}  
where the second and fourth contributions in Eq.~(\ref{tm1}) start
at order $\alpha_s$ since $\mathcal{L}_m^{(1)}$ contains at
least one collinear gluon. Note that all these terms are suppressed by
$\lambda^2$ compared to the leading contributions in $\rm{SCET_I}$. 
Only the third and fourth terms are nonzero due to the
spin structure of the currents. The mass term flips the spin of the
collinear quark, and therefore there must be even powers of $m$ to conserve
spin.  

\begin{figure}[t]
\begin{center}
\epsfig{file=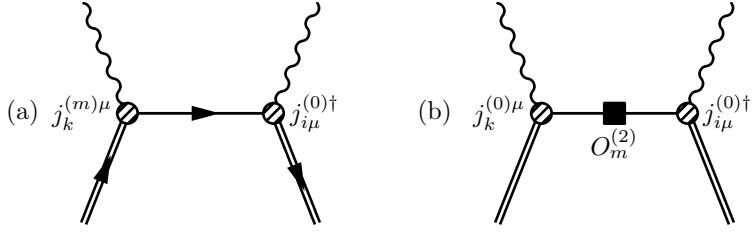, width=10cm}
\end{center}
\vspace{-0.5cm}
\caption{Tree-level Feynman diagrams for the leading 
mass corrections to $\overline{B}\rightarrow X_s \gamma$ near the
endpoint region in $\rm{SCET_I}$. The mirror image of the diagram (a)
should be included. The intermediate hard-collinear strange quark is
integrated out at $\mu\sim \sqrt{m_b \Lambda}$ to match onto
$\rm{SCET_{II}}$.}
\label{fig5}
\end{figure}

The tree-level Feynman diagrams for the mass corrections to
$\overline{B} \rightarrow X_s \gamma$ in the endpoint region are shown
in Fig.~\ref{fig5}. Fig.~\ref{fig5} (a) is zero as explained
above. Fig.~\ref{fig5} (b) yields 
\begin{equation} 
M_{ik,\mu\nu}^{(b)}= -i \int d^4 z  d^4 y~ 
e^{-i r\cdot z} T \Bigl[j_{i\mu}^{(0)\dagger}(z),
i\mathcal{L}_m^{(2)}(y), j_{k\nu}^{(0)}(0) \Bigr], 
\end{equation}
where the leading heavy-to-light currents $j_{k\mu}^{(0)}$ ($k=1,2$) are
given by
\begin{equation} \label{jmu}
j_{1\mu}^{(0)} = i \overline{\xi}_n W \gamma^{\perp}_{\mu}(1-\gamma_5)
h_v -\frac{i}{2} \overline{\xi}_n W  \overline{n}_{\mu}(1+\gamma_5)
h_v, \  j_{2\mu}^{(0)} = i \overline{\xi}_n W
\gamma^{\perp}_{\mu}(1-\gamma_5) h_v.   
\end{equation}
The amplitude $M_{ik,\mu\nu}^{(b)}$ is given as
\begin{eqnarray} \label{mik}
M_{ik,\mu\nu}^{(b)} &=& - \int d^4 z d^4 y  e^{-i r\cdot z}~ T \Bigl[
 j_{i\mu}^{(0)\dagger} (z) \overline{\xi}_n
 \frac{m^2}{\overline{n}\cdot p} \nn \xi_n (y) j_{k\nu}^{(0} (0)
 \Bigr]  \\ 
&=& \frac{m^2}{\overline{n}\cdot p} \int \frac{dn\cdot k
 d\overline{n}\cdot z}{4\pi} e^{-i n\cdot (r+k) \overline{n}\cdot z/2}
 \Bigl [J_P (n\cdot k) \Bigr]^2 T\Bigl[ \overline{h}_v
 Y\Bigl(\frac{\overline{n}\cdot z}{2} \Bigr) \Gamma_{ik,\mu\nu}^{(b)}
 Y^{\dagger}  h_v(0) \Bigr] \nonumber 
\end{eqnarray}
where the Dirac structure $\Gamma_{ik,\mu\nu}^{(b)}$ is given by
\begin{eqnarray} \label{gamik}
\Gamma_{11,\mu\nu}^{(b)} &=& \gamma^{\perp}_{\mu} \fms{n}
\gamma^{\perp}_{\nu} (1-\gamma_5) -\frac{\overline{n}_{\mu}}{2}
\fms{n} \gamma^{\perp}_{\nu} (1-\gamma_5)
-\frac{\overline{n}_{\nu}}{2} \gamma^{\perp}_{\mu} \fms{n}
(1+\gamma_5) +\frac{\overline{n}_{\mu} \overline{n}_{\nu}}{4} \fms{n}
(1+\gamma_5), \nonumber \\
\Gamma_{12,\mu\nu}^{(b)} &=& \gamma^{\perp}_{\mu} \fms{n}
\gamma^{\perp}_{\nu} (1-\gamma_5) -\frac{\overline{n}_{\mu}}{2}
\fms{n} \gamma^{\perp}_{\nu} (1-\gamma_5), \nonumber \\
\Gamma_{21,\mu\nu}^{(b)} &=& \gamma^{\perp}_{\mu} \fms{n}
\gamma^{\perp}_{\nu} (1-\gamma_5)-\frac{\overline{n}_{\nu}}{2}
\gamma^{\perp}_{\mu} \fms{n} (1+\gamma_5), \
\Gamma_{22,\mu\nu}^{(b)} = \gamma^{\perp}_{\mu} \fms{n}
\gamma^{\perp}_{\nu} (1-\gamma_5). 
\end{eqnarray}
In Eq.~(\ref{mik}), the ultrasoft interactions were decoupled from the
collinear field, and the resultant usoft Wilson line is given by 
\begin{equation}
Y (x) = \Bigl[\sum_{\rm{perms}} \exp \Bigl(-g 
\frac{1}{n \cdot \mathcal{P}}
n \cdot A_{us} (x) \Bigr) \Bigr],
\end{equation}
and $n \cdot \mathcal{P}$ is of order $\Lambda$. 
In obtaining Eq.~(\ref{mik}), we use the definition of the jet
function 
\begin{equation}
\langle 0| T\Bigl[ W^{\dagger} \xi_n (z) \overline{\xi}_n W \Bigr]
|0\rangle = i \frac{\fms{n}}{2}  \int \frac{d^4 k}{(2\pi)^4}
e^{-ik\cdot z} J_P (k),   
\end{equation}
where $P=\overline{n}\cdot p$ is the label momentum, and the jet
function is a function of $n\cdot k$ only with $J_P (k) =J_P (n\cdot
k)$. 

The matrix element of the remaining operators can be written as
\begin{eqnarray} \label{bmat}
&&\langle \overline{B}_v | \overline{h}_v
Y(\frac{\overline{n}\cdot z}{2})\Gamma_{ik,\mu\nu}^{(b)}
Y^{\dagger} h_v(0)| \overline{B}_v \rangle  \nonumber \\
&=& \int dn\cdot l e^{in\cdot l \overline{n} \cdot z/2} \langle
\overline{B}_v |\overline{h}_v  Y \Gamma_{ik,\mu\nu}^{(b)} \delta
(n\cdot l -n\cdot i\partial) Y^{\dagger}  h_v(0)| \overline{B}_v
\rangle \nonumber \\ 
&=& \int dn\cdot l e^{in\cdot l \overline{n} \cdot z/2} \mathrm{tr}
\Bigl( \frac{P_v}{2} \Gamma_{ik,\mu\nu}^{(b)} \Bigr)  \langle
\overline{B}_v |\overline{h}_v  Y \delta
(n\cdot l -n\cdot i\partial) Y^{\dagger}  h_v(0)| \overline{B}_v
\rangle \nonumber \\
&=& -2 (g^{\perp}_{\mu\nu} - i\epsilon^{\perp}_{\mu\nu}) \int dn\cdot
l e^{in\cdot l \overline{n} \cdot z/2} f^{(0)} (n\cdot l), 
\end{eqnarray}
where $P_v = (1+\fms{v})/2$ is the projection operator for the heavy
quark, and $f^{(0)}$ is the leading shape function of the $B$ meson,
which is defined as
\begin{eqnarray} 
f^{(0)}(n\cdot l)&=&\frac{1}{2} \int \frac{d\overline{n} \cdot z}{4\pi}  
e^{-i n\cdot l \frac{\overline{n} \cdot z}{2}} 
\langle \overline{B}_v |\overline{h}_v Y(\frac{\overline{n}\cdot
  z}{2})Y^{\dagger} h_v (0)| \overline{B}_v  \rangle \nonumber \\
&=& \frac{1}{2} 
~\langle \overline{B}_v |\overline{h}_v Y \delta(n\cdot l - n \cdot
i\partial) Y^{\dagger} h_v| \overline{B}_v \rangle,
\end{eqnarray} 
with $\epsilon^{\perp}_{\mu\nu} = \epsilon_{\mu\nu\alpha\beta}
\overline{n}^{\alpha} n^{\beta}/2$. Note that the final result of
Eq.~(\ref{bmat}) is independent of $\Gamma_{ik}^{(b)}$ since only the first term
in each Dirac structure in Eq.~(\ref{gamik}) contributes. 
 
The forward scattering amplitude with the mass correction at tree
level can be written as    
\begin{eqnarray} \label{factm}
T_{\mu\nu}^{(m)} (\omega) 
&=&-\tilde{H}(\omega, m_b, \mu_0) E_{\gamma}^2 (g^{\perp}_{\mu\nu}
-i\epsilon^{\perp}_{\mu\nu} ) \int dn\cdot k J_{\omega}^{(m)} (n\cdot
k) f^{(0)} \bigl(n\cdot k -m_b (1-\x) \bigr) \\
&=&- \tilde{H}(\omega, m_b, \mu_0) E_{\gamma}^2 (g^{\perp}_{\mu\nu}
-i\epsilon^{\perp}_{\mu\nu} ) \int dn\cdot l f^{(0)} (n\cdot l)
J_{\omega}^{(m)} (m_b (1-\x) +n\cdot l, \mu_0, \mu), \nonumber 
\label{fac1}
\end{eqnarray}
where $\mu_0$ is a typical scale where $\rm{SCET_{I}}$
can be matched onto $\rm{SCET_{II}}$, and 
$\omega$ is fixed as $\overline{n}\cdot p$ by the delta function 
in Eq.~(\ref{jmunu}). Here we use the relation $n\cdot l= n\cdot k +
n\cdot r = n\cdot k +\overline{\Lambda}-n\cdot p_X$.
$J_{\omega}^{(m)} (n\cdot k)$
is the jet function obtained from the matching between $\rm{SCET_I}$ and
$\rm{SCET_{II}}$, with the tree-level result given by 
\begin{equation}  
J_{\omega}^{(m)}(n\cdot k) |_{\rm{tree}} =  \frac{m^2}{\omega} \Bigl(
J_{\omega} (n\cdot k) \Bigr)^2 = \frac{m^2}{\omega}
\frac{1}{(n\cdot k + i \epsilon)^2}.
\label{jo}
\end{equation}
The hard factor $\tilde{H}(\omega,m_b,\mu_0)$ is obtained 
by matching the heavy-to-light currents between the full theory
and  $\rm{SCET_I}$ and is evolved to the scale $\mu \sim \mu_0$. At
$\mu = m_b$, it is given by 
\begin{equation} 
\tilde{H}(\omega,m_b,\mu=m_b) = |C_1(\omega,m_b) + C_2(\omega,m_b)|^2. 
\end{equation} 
Since the invariant mass of the jet is given by $p_X^2 \ge 0$, the
range of the residual momentum $n\cdot k$ is $0 \le n\cdot k \le
n\cdot p_X$. Also the residual momentum of the heavy quark $n\cdot l$
is smaller than the $B$ meson residual mass $\overline{\Lambda} = m_B
- m_b$.

To proceed further, there are two possible types of formulations. 
One, based on the second expression in Eq.~(\ref{factm}), is 
useful in the moment analysis and will be
illustrated in the next subsection. 
Here, we begin with the first expression in Eq.~(\ref{factm}), in
which we write $n\cdot l$ as \cite{Lee:2004ja}
\begin{equation} 
n\cdot l = \overline{\Lambda} - (1-z) n\cdot p_X,
\end{equation} 
where $z$ is given by 
\begin{equation} 
n\cdot k = z n\cdot p_X,~~~0 \le z \le 1.
\end{equation}
Then $T_{\mu\nu}^{(m)}$ can be written as 
\begin{eqnarray} \label{tmm}
T_{\mu\nu}^{(m)}&=& - E_{\gamma}^2
(g^{\perp}_{\mu\nu} -i\epsilon^{\perp}_{\mu\nu}
)\tilde{H}(\overline{n}\cdot p_X, m_b, \mu_0) n \cdot p_X  \\ 
&\times& \int_0^1 dz ~J_{\overline{n}\cdot p_X}^{(m)}(zn\cdot
p_X,\mu_0, \mu)  
~f^{(0)} (\overline{\Lambda}- (1-z) n\cdot p_X,\mu) \nonumber \\ 
\label{fac3}
&\equiv& - E_{\gamma}^2 (g^{\perp}_{\mu\nu} -i\epsilon^{\perp}_{\mu\nu}
)\tilde{H}(\overline{n}\cdot p_X, m_b, \mu_0) 
\int_0^1 d z ~\mathcal{J}_m(z,p_X^2, \mu) 
~f^{(0)} (\overline{\Lambda}- (1-z) n\cdot p_X, ~\mu), \nonumber
\end{eqnarray} 
where we use the relation $\omega=\overline{n}\cdot
p=\overline{n}\cdot p_X$  at leading order. The dimensionless jet
function $\mathcal{J}_m$ is given as 
\begin{eqnarray} 
\mathcal{J}_m (z, p_X^2, \mu) &=& n\cdot p_X~ 
J_{\overline{n}\cdot p_X}^{(m)}(zn\cdot p_X,\mu_0, \mu), 
\nonumber \\ 
\label{jm}
\mathcal{J}_m^{(0)} (z, p_X^2, \mu)
&=& \frac{m^2}{p_X^2}~ \frac{1}{(z+i\epsilon)^2}, 
\end{eqnarray}
where $\mathcal{J}_m^{(0)}$ is the jet function at tree level. 
When we take a discontinuity of the forward scattering amplitude, the
imaginary part entirely comes from the jet function, which is given by 
\begin{equation} 
\mathrm{Im}~\Bigl(~\frac{1}{\pi} \mathcal{J}_m^{(0)}~ \Bigr)
 = \frac{m^2}{p_X^2}~ 
\frac{d}{d z}~ \delta(z). 
\label{im}
\end{equation}

As shown in Eq.~(\ref{tmm}), the forward scattering amplitude can be
expressed as a convolution of the jet function and the shape function;
that is, it is given by the factorized form with the hard factor, the
jet function with the mass correction, and the leading-order shape
function. Note that the effect of the quark mass, even at higher orders
in $m^2/p_X^2$, resides only in the jet functions as it should, and
it does not affect the $B$-meson shape function. In order to obtain
the full subleading contributions to the decay rate, we should include
the mass correction in the jet functions, the contribution of the
subleading $B$-meson shape functions induced by the high-dimensional
heavy-quark bilinears, which can be studied in the framework of HQET,   
and the effects of the subleading heavy-to-light currents
in taking the time-ordered products. 

As can be seen in Eqs.~(\ref{jm}) and (\ref{im}), the subleading jet function
from the mass correction is suppressed by $m^2/p_X^2$ or $m^2/(m_b\Lambda)$
compared to the leading jet function. This contribution is of order
$\Lambda/m_b$ if we treat the strange quark mass to be of order
$\Lambda$. This is one of the examples in which the subleading terms
are formally suppressed by $\Lambda^2/m_b^2$, but they are actually
suppressed by $\Lambda/m_b$ near the endpoint. 

\begin{figure}[t]
\begin{center}
\epsfig{file=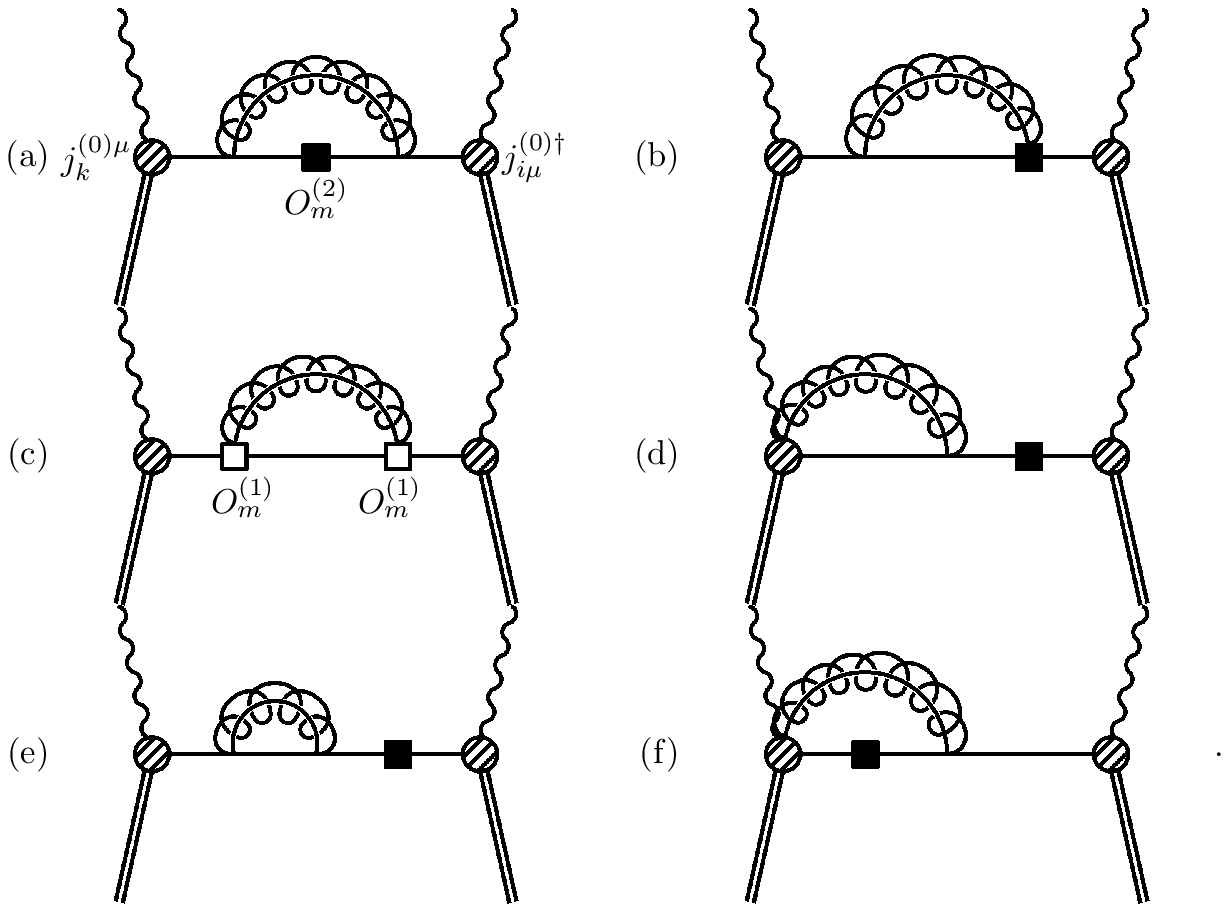, width=10cm}
\end{center}
\vspace{-0.5cm}
\caption{Feynman diagrams for the jet function at one loop. The mirror
  diagrams of (b), (d), (e), and (f) are omitted. The Sudakov
  double logarithms are produced  only in the diagram (d).}
\label{fig6}
\end{figure}

We can expand the jet functions in powers of $m^2/p_X^2$ and
$\alpha_s$. The jet function at first order in $m^2/p_X^2$ and in
$\alpha_s$ can be computed, with the
relevant Feynman diagrams shown in Fig.~\ref{fig6}. The contributions
of Fig.~\ref{fig6} (a), (b), (c) are obtained from Eq.~(\ref{O2off}),
and Fig.~\ref{fig6} (d), (e), (f) with their hermitian conjugates are
given as
\begin{eqnarray}
\sum_{i=a,b,c}\mathcal{J}_{m,i}^{(1)} &=& \frac{\alpha_s C_F}{4\pi}
\frac{m^2}{p_X^2} \frac{1}{(z+i\eps)^2} (-p_X^2)^{-\eps}
\Bigl(\frac{7}{\eps} +9 \Bigr), \nonumber \\
\mathcal{J}_{m,d}^{(1)} &=&  \frac{\alpha_s C_F}{4\pi}
\frac{m^2}{p_X^2} \frac{1}{(z+i\eps)^2} (-p_X^2)^{-\eps}
\Bigl(\frac{4}{\eps^2} +\frac{4}{\eps} +8-\frac{\pi^2}{3} \Bigr),
\nonumber \\ 
\mathcal{J}_{m,e}^{(1)} &=&  \frac{\alpha_s C_F}{4\pi}
\frac{m^2}{p_X^2} \frac{1}{(z+i\eps)^2} (-p_X^2)^{-\eps}
\Bigl(-\frac{2}{\eps} -2 \Bigr), \nonumber \\
\mathcal{J}_{m,e}^{(1)} &=&  \frac{\alpha_s C_F}{4\pi}
\frac{m^2}{p_X^2} \frac{1}{(z+i\eps)^2} (-p_X^2)^{-\eps}
\frac{4}{\eps}. 
\end{eqnarray}
Therefore the jet function at order
$\alpha_s$ is given as 
\begin{eqnarray} 
\mathcal{J}_m^{(1)} (z, p_X^2, \mu) &=& \frac{\alpha_s C_F}{4\pi}
\frac{m^2}{p_X^2} \frac{1}{(z+i\epsilon)^2}
\Biggl[2\Bigl(\ln \frac{p_X^2}{\mu^2} + \ln (-z-i\epsilon) \Bigr)^2 
\nonumber \\
&& -13\Bigl(\ln \frac{p_X^2}{\mu^2} + \ln (-z-i\epsilon) \Bigr) 
+15 - \frac{\pi^2}{3} \Biggr].
\end{eqnarray}
The discontinuity of the jet function to
order $\alpha_s$ is given by
\begin{eqnarray} 
\mathrm{Im}~\Bigl(\frac{1}{\pi} \mathcal{J}_m~ \Bigr)
&=&\frac{m^2}{p_X^2}\frac{d}{dz} 
\Biggl{\{} \delta (z) + \frac{\alpha_s}{4\pi}C_F 
\Biggl[\Bigl( 2\ln^2 \frac{p_X^2}{\mu^2}-9\ln\frac{p_X^2}{\mu^2} 
+ 10-\pi^2 \Bigr) \delta (z) \nonumber \\ 
&&+\Bigl(4\ln\frac{p_X^2}{\mu^2} -9\Bigr) \frac{1}{(z)_+}
+4 \Biggl(\frac{\ln(z)}{z}\Biggr)_+\Biggr] \Biggr{\}}.  
\label{img}
\end{eqnarray}

\subsection{Moment analysis of the mass correction}
In the endpoint region $\x \to 1$ ($\x = 2E_{\gamma}/m_b$), it is
useful to consider the moments of the differential decay rate and take
the large $N$ limit. In the limit $1-\x \sim 1/N \sim \Lambda/m_b$, 
the moments of the mass correction to the differential decay rate 
can be studied. In doing the moment analysis, it is convenient to
employ the second expression for $T^{(m)}_{\mu\nu}$ in
Eq.~(\ref{factm}).  Consider the quantity
\begin{equation} 
S=\int dn\cdot l f^{(0)} (n\cdot l) J_{\omega}^{(m)} (m_b (1-\x)
+n\cdot l, \mu_0, \mu),
\label{tm}
\end{equation}
with the momentum of the strange quark $p_s^{\mu} = m_b n^{\mu}/2 +
k^{\mu} + m_b (1-\x) \overline{n}^{\mu} /2$. 
The jet function has support for $-m_b(1-\x) \le n\cdot l 
\le \overline{\Lambda}$, and if we let $n\cdot l =-(1-y) m_b$ 
Eq.~(\ref{tm}) can be written as 
\begin{equation} 
S= m_b \int_{\x}^{m_B/m_b} dy f^{(0)} \bigl(-m_b (1-y) \bigr)
J^{(m)}_{m_b} \bigl(y,z_{\gamma}\bigr), 
\end{equation}
where the jet function proportional to $m^2$ to order $\alpha_s$ is
given by
\begin{eqnarray}
J_{\omega}^{(m)} (y,z_{\gamma})&=& \frac{m^2}{\omega m_b^2 y^2}
\frac{1}{(1-z_{\gamma}+i\epsilon)^2} \\
&\times& \Bigl[ 1+\frac{\alpha_s C_F}{4\pi} \Bigl( 2\ln^2
\frac{-\omega m_b y (1-z_{\gamma})}{\mu^2} -13 \ln  \frac{-\omega m_b
  y   (1-z_{\gamma})}{\mu^2} +15-\frac{\pi^2}{3} \Bigr) \Bigr],
\nonumber  
\end{eqnarray}
where $z_{\gamma}= \x/y$. Taking the imaginary part, we obtain
\begin{eqnarray}
-\frac{1}{\pi} \mathrm{Im}
J^{(m)}_{m_b} (y,z_{\gamma}) 
&=&\frac{m^2}{m_b^3 y^2} \Bigl\{  \Bigl[ 1 + \frac{\alpha_s
  C_F}{4\pi} \Bigl( 2\ln^2 \frac{m_b^2 y}{\mu^2}  -9 \ln \frac{m_b^2
  y}{\mu^2} +10 -\pi^2 \Bigr) \Bigr] \frac{d}{dz_{\gamma}} \delta
(1-z_{\gamma}) \nonumber \\ 
&+&\frac{\alpha_s C_F}{4\pi} \Bigl[  \Bigl( 4   \ln \frac{m_b^2
  y}{\mu^2} -9\Bigr) \frac{d}{dz_{\gamma}} \frac{1}{(1-z_{\gamma})_+}
+4 \frac{d}{dz_{\gamma}} \Bigl(\frac{\ln
  (1-z_{\gamma})}{1-z_{\gamma}} \Bigr)_+ \Bigr] \Bigr\} \nonumber \\   
&\equiv& \frac{1}{m_b y^2} \overline{\mathcal{J}}^{(m)}
(z_{\gamma}), 
\end{eqnarray}
where we neglect $\ln y$ terms near the endpoint $y\rightarrow 1$, and
$\overline{\mathcal{J}}^{(m)} (z_{\gamma})$ is a dimensionless function
  suppressed by $m^2/m_b^2$. 

Finally the mass correction to the differential decay rate can be
written as   
\begin{equation} 
\frac{1}{\Gamma_0}\frac{d \Gamma^{(m)}}{d\x}
=  \tilde{H}(m_b, \mu_0)
\x^3\int^1_{\x}  \frac{dy}{y^2} \overline{f}^{(0)}(y, \mu)
\overline{\mathcal{J}}^{(m)}\Bigl(\frac{\x}{y},\mu_0, \mu \Bigr),
\end{equation} 
where $\overline{f}^{(0)}(y,\mu) = f^{(0)}(-(1-y)m_b,\mu)/m_b$, and the 
difference between $m_b$ and $m_B$ is neglected because it is
subleading. The moments of the mass correction to the differential
decay rate are given by 
\begin{eqnarray} \label{moments} 
\int^1_{0} d\x \x^{N-1} \frac{1}{\Gamma_0}\frac{d \Gamma^{(m)}}{d\x}
&=& \tilde{H}(m_b, \mu_0) \int_0^1 d\x \x^{N+2}
\int^1_{\x} \frac{dy}{y^2} \overline{f}^{(0)}(y,\mu) 
\overline{\mathcal{J}}^{(m)}\bigl(\frac{\x}{y},\mu_0,\mu \bigr)
\nonumber \\ 
&=&  \tilde{H}(m_b, \mu_0) \int_0^1 dy
y^{N+1} \overline{f}^{(0)} (y) \int_0^1 dz_{\gamma} z_{\gamma}^{N+2}
\overline{\mathcal{J}}^{(m)} (z_{\gamma}).
\end{eqnarray} 
Therefore the moment of the differential decay rate is given by the
product of the moments of the shape function and the moments of the jet 
function. The moments $\overline{\mathcal{J}}^{(m)}_N$ to order
$\alpha_s$ are given by 
\begin{eqnarray} 
\overline{\mathcal{J}}^{(m)}_N &=& \int_0^1 dz_{\gamma} z_{\gamma}^{N-1}
\overline{\mathcal{J}}^{(m)} (z_{\gamma}) \nonumber \\
&=& -\frac{m^2}{m_b^2}(N-1) \Bigl[ 1+\frac{\alpha_s C_F}{4\pi} \Bigl( 2\ln^2
\frac{m_b^2}{\mu^2} +(4H_{N-2} -9) \ln \frac{m_b^2}{\mu^2}
\nonumber \\
&&+4\sum_{j=1}^{N-2} \frac{H_j}{j} -9H_{N-2} +10-\pi^2 \Bigr) \Bigr],  
\end{eqnarray} 
where $\ln y$ is neglected in the limit $y\rightarrow 1$, and $H_j =
\sum_{k=1}^j 1/k$. In the large $N$ limit, this becomes
\begin{equation} 
\overline{\mathcal{J}}^{(m)}_N (\mu) = - \frac{m^2}{m_b^2}N
\Biggl[ 1 + \frac{\alpha_s}{4\pi} C_F \Bigl( 2 \ln^2
\frac{m_b^2}{\mu^2\overline{N}} - 9 \ln\frac{m_b^2}{\mu^2\overline{N}}
+10-\frac{2\pi^2}{3}  \Bigr)\Biggr],
\label{imjet} 
\end{equation}
where $\overline{N} = N e^{\gamma_E}$. 

For comparison, the leading result without the quark mass term can be
written in the same way as Eq.~(\ref{factm}) with the jet function replaced by
\begin{eqnarray}
 J_{\omega}^{(0)} \bigl(y,z_{\gamma} \bigr)  &=&
 \frac{1}{m_b y(1-z_{\gamma} +i\epsilon)} \Bigl[ 1+\frac{\alpha_s
 C_F}{4\pi}   \Bigl( 2\ln^2 \frac{-\omega m_b y (1-z_{\gamma}
 +i\epsilon)}{\mu^2}  \nonumber   \\
&&-3 \ln  \frac{-\omega m_b y
 (1-z_{\gamma}+i\epsilon)}{\mu^2}+7-\frac{\pi^2}{3}   \Bigr) \Bigr],
\end{eqnarray}
with discontinuity 
\begin{eqnarray} \label{lj0}
-\frac{1}{\pi} \mathrm{Im}
J_{m_b}^{(0)} (y, z_{\gamma})
&=& \frac{1}{m_b y} \Bigl\{ \delta
(1-z_{\gamma}) \Bigl[ 1+\frac{\alpha_s C_F}{4\pi} \Bigl( 2\ln^2
\frac{m_b^2   y}{\mu^2} -3 \ln \frac{m_b^2 y}{\mu^2} +7-\pi^2 \Bigr)
\Bigr] \nonumber \\   
&+& \frac{\alpha_s C_F}{4\pi} \Bigl[ 4 \Bigl( \frac{\ln
  (1-z_{\gamma})}{1-z_{\gamma}} 
\Bigr)_+ +\Bigl( 4\ln  \frac{m_b^2 y}{\mu^2} -3 \Bigr)
\frac{1}{(1-z_{\gamma})_+} \Bigr] \Bigr\} \theta(z_{\gamma}) \theta
(1-z_{\gamma}) \nonumber \\ 
&\equiv& \frac{1}{m_b y} \overline{\mathcal{J}}^{(0)} (z_{\gamma},
\mu),  
\end{eqnarray}
where we again neglect $\ln y$ terms near the endpoint $y\rightarrow
1$.  This is consistent with the result in Ref.~\cite{Lee:2004ja}.

The leading differential decay rate is given by
\begin{equation}
\frac{1}{\Gamma_0} \frac{d\Gamma}{d\x} = \x^3 \Bigl(\frac{m_b}{m_B}
\Bigr)^3 \tilde{H} (m_b,\mu_0) \int_{\x}^{m_B/m_b} \frac{dy}{y}
\overline{f}^{(0)} (y,\mu) \overline{\mathcal{J}}^{(0)}
\bigl(\frac{\x}{y}, \mu_0, \mu \bigr),  
\end{equation}
with moments
\begin{eqnarray}
&&\int_0^{m_B/m_b} d\x \x^{N-1} \frac{1}{\Gamma_0} \frac{d\Gamma}{d\x}
= \Bigl( \frac{m_B}{m_b} \Bigr)^3 \tilde{H} (m_b, \mu_0)
\int_0^{m_B/m_b} d\x \x^{N+1} \nonumber \\
&&\times \int_{\x}^{m_B/m_b} \frac{dy}{y} \overline{f}^{(0)} (y,\mu )
\overline{\mathcal{J}}^{(0)} \bigl( \frac{\x}{y}, \mu_0, \mu \bigr)
\nonumber \\
&=& \Bigl(\frac{m_B}{m_b} \Bigr)^3 \int_{0}^{m_B/m_b} dy y^{N+2}
\overline{f}^{(0)} (y,\mu ) \int_0^1 dz_{\gamma} z_{\gamma}^{N+2}
\overline{\mathcal{J}}^{(0)} (z_{\gamma}, \mu_0, \mu).
 \end{eqnarray} 
The moments of the differential decay rate is factorized into
a product of the moments of the jet function and the moments of the
shape function of
the $B$ meson. The moments of the leading-order jet function become
\begin{eqnarray}
\overline{\mathcal{J}}^{(0)}_N &=& \int_0^1 dz_{\gamma}
z_{\gamma}^{N-1} \overline{\mathcal{J}}^{(0)} (z_{\gamma}) \nonumber \\
&=&  1+\frac{\alpha_s C_F}{4\pi} \Bigl[ 2\ln^2 \frac{m_b^2}{\mu^2}
-(4H_{N-1} +3) \ln \frac{m_b^2}{\mu^2} +3H_{N-1} 
+4 \sum_{j=1}^{N-1} \frac{H_j}{j} +7-\pi^2 \Bigr] \nonumber \\
&\longrightarrow& 1+ \frac{\alpha_s C_F}{4\pi} \Bigl( 2\ln^2
\frac{m_b^2}{\mu^2 \overline{N}} -3 \ln \frac{m_b^2}{\mu^2
\overline{N}} +7-\frac{2\pi^2}{3}  \Bigr).
\end{eqnarray}

In Eq.~(\ref{moments}), the typical matching scale $\mu_0$ between 
$\rm{SCET_I}$ and $\rm{SCET_{II}}$ can be considered as 
$m_b/\sqrt{\overline{N}}$ 
and the hard coefficients $\tilde{H}$ evolves 
from the scale $\mu = m_b$ to $m_b/\sqrt{\overline{N}}$.  The jet
function, determined at the scale $\mu=m_b/\sqrt{\overline{N}}$, 
scales down to the arbitrary scale $\mu < m_b/\sqrt{\overline{N}}$.  
Since the product of $\overline{f}^{(0)}_N(\mu)$ and 
$\overline{\mathcal{J}}^{(m)}_N (\mu_0, \mu)$ 
is independent of the renormalization scale $\mu$, 
the renormalization behavior of $\overline{\mathcal{J}}^{(m)}_N (\mu_0, \mu)$ 
is easily determined from 
the scaling of $\overline{f}^{(0)}_N(\mu)$, 
which is given as \cite{Bauer:2000ew,Inclusive}
\begin{equation} 
\mu \frac{d}{d\mu} \overline{f}^{(0)}_N(\mu) = - \gamma_N (\mu)~ 
\overline{f}^{(0)}_N(\mu), 
\end{equation} 
where $\gamma_N$ is the anomalous dimension of the shape function. For
large $N$ and at order $\alpha_s$, it is given by
\begin{equation} 
\gamma_N(\mu) = - \frac{\alpha_s C_F}{\pi} \Bigl(1 + 
2\ln\frac{m_b}{\mu \overline{N}} \Bigl).
\end{equation}  

Finally, since the hard part and the shape function in the leading
mass correction of the moments of $\overline{B} \to X_s \gamma$ are
the same as those of the leading-order moments \cite{Bauer:2000ew}, we
find the resummation for the moments of the leading mass correction at
the scale $\mu = m_b/\overline{N}$ to order $\alpha_s$ can be written
as    
\begin{eqnarray} 
\int^1_{0} \x^{N-1} \frac{1}{\Gamma_0}\frac{d \Gamma^{(m)}}{d\x} 
\Bigl(\frac{m_b}{\overline{N}} \Bigr)
&=& \overline{f}^{(0)}_N \Bigl( \frac{m_b}{\overline{N}} \Bigr)
\Biggl(\frac{\alpha_s (m_b/\sqrt{\overline{N}})}{\alpha_s(m_b)}
\Biggr)^{\frac{C_F}{\beta_0}\Bigl(\frac{5-8\pi}{\beta_0 \alpha_s}\Bigr)}
\Biggl(-N \frac{m^2(m_b/\sqrt{\overline{N}})}{m_b^2} \Biggr) \nonumber
\\  
\label{LO} 
&& \times 
\Biggl(\frac{\alpha_s(m_b/\overline{N})}{\alpha_s (m_b/\sqrt{\overline{N}})}
\Biggr)^{\frac{2C_F}{\beta_0}
\Bigl(1+ \frac{4\pi}{\beta_0 \alpha_s}-2\ln\overline{N} \Bigr)}.
\end{eqnarray}
This result represents that the leading mass corrections are of order
$(m^2/m_b^2) N \ln^k N $ in the large $N$ limit, and they are 
resummed in  moment space. Compared with the leading-order 
moments, the leading mass correction is always suppressed by 
$N m^2/m_b^2  \sim \Lambda/m_b$. 

A note is in order for Eq.~(\ref{LO}), in which the quark mass $m$ is
evaluated at $m_b/\sqrt{\overline{N}}$ instead of
$m_b/\overline{N}$. This means that the strange quark mass is frozen at
$\mu = m_b/\sqrt{\overline{N}}$, and the scaling behavior below that
scale to $m_b/\overline{N}$ resides in the jet function. This is
motivated by the fact that the effects of the quark mass reside only
in the jet function and it looks more transparent to consider the
scaling behavior of the jet function as a whole including the quark
mass. As can be seen in Fig.~\ref{fig6} (a)-(c), the radiative
correction for the quark mass is included in computing the jet
function. Another equivalent way of expressing Eq.~(\ref{LO}) is to
separate the effects of the quark mass from the remainder of the jet
function and to scale each of them. Then the result can be expressed
with $m^2$ evaluated at $m_b/\overline{N}$. These two methods are
equivalent, and the latter method may be
useful in considering the effects of the quark mass in exclusive $B$
decays.

From this analysis we find that the quark mass effect to the leading
decay rate is of order $Nm^2/m_b^2 \sim m^2 /m_b^2 (1-x_{\gamma}) \sim
m^2 /m_b \Lambda$  near the endpoint region
$1-x_{\gamma} \sim \Lambda/m_b$. The size of the subleading
corrections in Ref.~\cite{Lee:2004ja} is of order $\Lambda/m_b$. The
quark mass corrections and the subleading corrections are of the same
order if we regard the strange quark mass as of order $m\sim
\Lambda$. However, the strange quark mass is numerically about 
80--130 MeV. Taking $\Lambda \sim $ 500 MeV, the quark mass correction
is about 2--7\% of other subleading corrections, and less than 1\%
compared to the leading decay rate. Therefore the mass effect can be
regarded as small compared to other subleading corrections, but the
important point is that the effect of the light quark mass can be
systematically implemented in the theoretical framework of SCET, and
as experimental uncertainties become smaller, this effect should also
be included.

\section{Conclusion} 
We have considered the contribution of a quark mass of order
$\Lambda$ in SCET and its application to $\overline{B} \rightarrow
X_s \gamma$ in the endpoint region. The quark mass can be included in the
SCET Lagrangian systematically by integrating out hard degrees of freedom. 
We can find an extended reparameterization invariance including the
quark mass, in which we modify only the reparameterization transformation
of the spinor for the transformation of type-II. As a result, the SCET
Lagrangian can be separated into two reparameterization-invariant
combinations. The subleading operators in each combination are
related to the leading operators in that combination by the
reparameterization invariance, and they have the same Wilson
coefficients as those of the leading operators. In particular, we find
that the mass operators in the SCET Lagrangian have  
trivial Wilson coefficients and are not renormalized.
These results are explicitly confirmed by the calculation of 
the corrections to the mass 
operators in SCET to one loop.  The extended reparameterization
invariance also constrains some of the Wilson coefficients for 
the heavy-to-light current operators with the quark mass. It plays  
an important role in the matching process of the subleading
heavy-to-light currents and the higher-order calculations of the
time-ordered products of the mass operators.

When we consider $\overline{B} \to X_s \gamma$ in the endpoint region, 
treating the strange quark mass to be of order $\Lambda$, the
subleading contribution is of order $\Lambda/m_b$. We have verified this
by matching the heavy-to-light current onto $\rm{SCET_{I}}$
with the mass operators. Many of the currents with the mass 
are related to the leading-order currents by the extended
reparameterization symmetry. There are also subleading operators
which are independent of the leading current, and these are obtained
at higher orders in $\alpha_s$. There are no contributions with odd
powers of $m$ to the decay rate because of spin conservation. 
The subleading contributions of order $m^2/p_X^2 \sim \Lambda/m_b$ 
come from the time-ordered products of the double spin-flipped
currents with the leading heavy-to-light currents, and the double
spin-flipped currents are obtained by the time-ordered products of the
leading currents with the mass operators in the SCET Lagrangian.   
The mass correction to the forward scattering amplitude 
is given by the factorized form which is expressed as a
convolution of the $m^2/p_X^2$ suppressed jet function and the leading 
shape function of the $B$ meson. The jet functions which are obtained
from the matching between $\rm{SCET_{I}}$ and $\rm{SCET_{II}}$ can be
always expanded by $m^2/p_X^2$, and can be computed perturbatively in
$\alpha_s$.

In some $B$ decays, the subleading effects can be important to extract
the CKM matrix elements. We have shown that the strange quark 
mass corrections give nonnegligible contributions of order
$\Lambda/m_b$ in $\bar{B} \to X_s \gamma$, and it would be interesting
to see if the mass corrections can give significant contribution to
other $B$ decays. The results in this paper can be a basis on how to
explain the $SU(3)$ flavor-symmetry breaking effects, as in
$B\rightarrow K^* \gamma$ and $B\rightarrow \rho \gamma$, in which the
mass effects could be a leading result.  

\section*{Acknowledgments}
J.~Chay was supported by Grant No. R01-2002-000-00291-0 from the Basic
Research Program of the Korea Science \& Engineering Foundation
and by Korea University. C.~Kim and A~.K.~Leibovich were supported by
the National Science Foundation under Grant No. PHY-0244599.  A.~K.~Leibovich was also supported in part by the Ralph E.~Powe Junior Faculty Enhancement Award.

\end{document}